\documentclass[10pt,twocolumn]{article}
\usepackage[margin=0.7in,columnsep=0.22in]{geometry}
\usepackage{amsmath,amssymb,graphicx,xcolor}
\usepackage[colorlinks=true,linkcolor=blue,citecolor=blue]{hyperref}
\newcommand{\p}{\partial}
\usepackage[T1]{fontenc}
\usepackage{tgpagella}
\usepackage{comment}
\usepackage{cite}
\usepackage{tikz}
\usetikzlibrary{arrows.meta,decorations.pathmorphing,calc}
\title{\vspace{-2em}\LARGE Pohlmeyer reduction of Newton-Cartan strings}
\author{Aritra Banerjee$^{*}$\\[2pt]\small Department of Physics, Birla Institute of Technology and Science - Pilani, Rajasthan, 333031, India}
\date{\vspace{-2em}}
\begin{document}
\twocolumn[{%
\maketitle
\begin{center}
\begin{minipage}{0.85\textwidth}
\textbf{Abstract.} We develop a Pohlmeyer-type reduction for classical non-relativistic strings on string Newton-Cartan backgrounds. For strings on a two-sphere, we identify the invariant reduced field, show that its dynamics is integrable, and construct a Lax representation. We relate this intrinsic formulation to the non-relativistic expansion of relativistic strings and obtain a dictionary between non-relativistic string solutions and their reduced counterparts. The construction explicitly determines which relativistic solitonic configurations survive the non-relativistic limit and which decouple. Moreover, it provides a framework for applying integrability methods to Newton-Cartan string dynamics.
\end{minipage}
\end{center}
\vspace{0.8em}}]
{\renewcommand\thefootnote{*}\footnotetext{\href{mailto:aritra.banerjee@pilani.bits-pilani.ac.in}{aritra.banerjee@pilani.bits-pilani.ac.in}}}

\section{Introduction}
Non-Lorentzian physics, and Galilean physics in particular, describes most of our daily experiences with nature. In high energy physics, by contrast, the most natural setting is the relativistic one. Recently, there has been renewed interest in non-relativistic (NR) field and string theories, propelled by advances in non-Lorentzian holography \cite{primer,Fontanella:2024rvn,Fontanella:2024kyl}. In the NR regime $c\to \infty$, the covariant formulation of such theories is based on Newton-Cartan manifolds \cite{cartan1986manifolds, HAVAS:1964vek}. The elegant geometric formulation gives a natural language to NR versions of field and string theories. In the current work, we focus on NR string theories \cite{GO,Danielsson:2000gi, Klebanov:2000pp, Danielsson:2000mu} in stringy Newton-Cartan (sNC) geometries \cite{Andringa:2012uz,BGY,Bergshoeff:2018vfn, Kluson:2018egd, Bergshoeff:2021bmc, Yan:2021lbe, BGRSY, Gomis:2019zyu}, see \cite{Fontanella:2026gaq} for a recent review. This particular setting singles out the time and one of the space directions such that the resulting non-Lorentzian spacetime splits into two longitudinal and $(D-2)$ transverse directions.\footnote{For the related torsional Newton-Cartan (tNC) geometry, see, for example, \cite{Christensen:2013lma, Christensen:2013rfa,Harmark1}.} String theory in this setting has been considered along many different directions, and the reader can see \cite{Relating,Bidussi,HH1,HH2,RoyRS2,Roy1,Roy2,FN2,Kluson,RoyLax,Fontanella:2020eje,Fontanella:2022fjd,Fontanella:2022pbm,Fontanella:2022wfj,Argandona:2025jhg} for these explorations, with the list nowhere being exhaustive.

However, there are some sNC string setups that have been less explored, thereby prompting the current work. Classical strings on symmetric spaces admit a reduction, due to Pohlmeyer \cite{Pohlmeyer}, in which the embedding is traded for invariant fields obeying an integrable equation of sine-Gordon type \cite{Larsen:1996gn, Miramontes}. For strings on $\mathbb R\times S^2$ the relativistic reduced theory is the sine-Gordon equation itself, whose elliptic solutions describe the BMN particle, the GKP string, giant magnons and spiky strings \cite{Plefka:2005bk} in a unified manner \cite{KMP1}. Its dressing generates new solutions, including those that realise the instabilities of their elliptic seeds \cite{KMP2,KMP3,KMP4}. Rigidly rotating and pulsating strings on spheres reduce, alternatively, to the integrable Neumann-Rosochatius system \cite{AFRT}. Our main goal, rooted in the discussion of integrability in non-Lorentzian sigma models, is to formulate a Pohlmeyer-equivalent reduction for the sNC strings, focusing on the case of $\mathbb R\times S^2$. We will discuss this in two ways, using an intrinsically NR sNC formalism, and via a consistent order-by-order expansion of the relativistic setup around $c\to\infty$.\footnote{Similar small $c$ string expansions can be performed for the opposite limit, for the \textit{Carrollian} target spaces, and more details can be found in \cite{Bagchi:2023cfp, Bagchi:2024rje}.}

For the intrinsic case, we will construct the Pohlmeyer field using the same geometric intuition behind its relativistic cousin: that of an invariant oriented area. In the sNC theory the Virasoro constraints fix only the longitudinal projections of the tangents to the string, and the natural invariant is the oriented area spanned by the tangents in the plane of the spatial longitudinal and transverse directions, normalised by their longitudinal projections. This field measures the transverse velocity of the string, and it obeys a tachyonic Klein-Gordon equation, which must be supplemented with appropriate boundary conditions. Furthermore, we show that its Lax pair is the Maurer-Cartan form of an intrinsic moving frame on the sNC sphere. In the expansion approach, the leading-order Virasoro constraint admits two inequivalent branches, which place the expansion at the two vacua of the sine-Gordon equation. On the \textit{winding} branch, the string is stretched along a meridian and almost at rest, near the unstable vacuum, and this branch reproduces the intrinsic theory. On the \textit{momentum} branch, the string is point-like and moves at nearly the speed of light, near the stable vacuum, and it has no intrinsic description. The Pohlmeyer field is a next-to-leading-order (NLO) object of the string expansion. Further, the expansion of the sine-Gordon Lax pair introduces an intriguing map to the intrinsic one.

The NR Pohlmeyer reduction also determines which relativistic solitons survive the NR limit. Rigidly rotating and pulsating strings appear as elliptic solutions in the vacuum corner of the moduli space, and small breathers on the line reduce to the pulsating string. Kinks, giant magnons \cite{HM} and spikes \cite{Kruczenski:2004wg, Kruczenski:2006pk} in static gauge, on the other hand, carry energies of order $c^2$ and decouple from the NR expansion.

The paper is organised as follows. In section~\ref{sec:pohl} we review the Pohlmeyer reduction of relativistic strings on $\mathbb R\times S^2$, and in section~\ref{sec:NC} we recall the sNC sigma model on this background. Section~\ref{sec:intrinsic} develops the intrinsic reduction: the Pohlmeyer field, its equation and boundary conditions, the intrinsic Lax pair and its monodromy, and the explicit map between strings and reduced fields. In section~\ref{sec:expansion} we turn to the $1/c^2$ expansion, discuss the winding and momentum sectors and their duality, and expand the Lax pair and its curvature order by order. Section~\ref{sec:kinks} explains why kinks decouple and how small breathers survive. We conclude with a summary of the novel elements and a discussion of future directions. The appendices collect more discussions and computations.

\section{Pohlmeyer reduction: Relativistic sphere}\label{sec:pohl}
Throughout this paper we use the light-cone coordinates $\sigma^\pm=\tau\pm\sigma$ and the derivatives $\p_\pm=\frac12(\p_\tau\pm\p_\sigma)$, so that $\p_+\p_-=\frac14(\p_\tau^2-\p_\sigma^2)$.\footnote{ These differ by factors of $\pm2$ from the derivatives used in \cite{KMP1}.} For the relativistic sigma model, the sphere of radius $R$ is embedded in $\mathbb R^{(1,3)}$, with time $X^0=f_+(\sigma^+)+f_-(\sigma^-)$. The Virasoro constraints fix the lengths $|\p_\pm\vec X|=|\p_\pm f_\pm|=|f_\pm'|$, and the only unconstrained quantity is the angle $\psi$ between the light-cone tangents,
\begin{equation}\label{eq:psidef}
\p_+\vec X\cdot\p_-\vec X=f_+'f_-'\cos\psi .
\end{equation}
We choose the linear gauge $f_\pm=m_\pm\sigma^\pm$ with $m_+m_->0$. 
Compatibility of the moving frame on the sphere built from $\vec X$ and $\p_\pm\vec X$ gives:
\begin{equation}\label{eq:SG}
\p_+\p_-\psi=-\frac{\mu^2}{4}\sin\psi,\qquad \mu^2=\frac{4m_+m_-}{R^2}.
\end{equation}
Further, in the static gauge, putting in $X^0=R\mu\tau$, i.e.\ $m_\pm=R\mu/2$, this becomes the sine-Gordon form: $\ddot\psi-\psi''+\mu^2\sin\psi=0$, and the constraints take the form \cite{KMP1}
\begin{equation}\label{eq:static}
|\p_\tau\vec X|^2=R^2\mu^2\cos^2\frac\psi2,\qquad |\p_\sigma\vec X|^2=R^2\mu^2\sin^2\frac\psi2 .
\end{equation}
Hence at the stable vacuum $\psi=0$,
$|\dot{\vec X}|=R\mu, |\vec X'|=0,$ the string moves at the speed of light and is point-like, while at the unstable vacuum $\psi=\pi$, $|\dot{\vec X}|=0, |\vec X'|=R\mu,
$ it is at rest and maximally stretched. The vacua of the sine-Gordon equation and their fluctuations are summarised in appendix~\ref{app:vacua}. The equation \eqref{eq:SG} is the flatness condition \(F_{+-}=\partial_- A_+-\partial_+ A_-+[A_+,A_-]=0\)
 of the $su(2)$-type Lax pair \cite{FT}
\begin{equation}\label{eq:Lax}
\begin{aligned}
A_+&=-\frac i2\,\p_+\psi\,\sigma_3+\frac{\lambda\mu}{4}\,\sigma_1,\\
A_-&=-\frac{\mu}{4\lambda}\big(\cos\psi\,\sigma_1+\sin\psi\,\sigma_2\big),
\end{aligned}
\end{equation}
whose curvature is $-\frac i2\big(\p_+\p_-\psi+\frac{\mu^2}{4}\sin\psi\big)\sigma_3$. Since the sine-Gordon equation \eqref{eq:SG} is integrable, the reduction makes the classical integrability of the underlying sigma model manifest.

\section{Newton-Cartan strings}\label{sec:NC}

Our focus in this work is on NR strings which, as discussed before, propagate on string Newton-Cartan (sNC) backgrounds with two longitudinal directions, and which arise either as an intrinsic sigma model or from the $1/c^2$ expansion of the relativistic string. The details from string sigma model calculations needed here \cite{Banerjee:2025gyh} are collected in appendix~\ref{app:NC}, and we quote the main results in the text whenever necessary. For $\mathbb R\times S^2$, with time $t$ and polar angle $\vartheta$ longitudinal and azimuth $\varphi$ transverse, the sNC data are
\begin{equation}\label{eq:sncdata}
\tau^0=dt,\quad \tau^1=d\vartheta,\quad E^2=\sin\vartheta\,d\varphi,\quad m^a=0 .
\end{equation}
The background \eqref{eq:sncdata} is torsion-free with $m^a=0$, so neither torsion nor Galilean boosts enter the reduction. For backgrounds whose longitudinal metric is not flat, such as the $AdS_2$ longitudinal geometry of non-relativistic $AdS_5\times S^5$ \cite{FN2}, the normalisation of the Pohlmeyer field introduced below acquires the longitudinal conformal factor, and the reduction must be modified accordingly.
In conformal gauge, with $X^\pm=t\pm\vartheta$, the intrinsic Lagrangian is
\begin{equation}\label{eq:NCL}
\mathcal L=2T\sin^2\!\vartheta\,\p_+\varphi\,\p_-\varphi+T\big(\xi\,\p_-X^+-\bar\xi\,\p_+X^-\big),
\end{equation}
where $\xi$ and $\bar\xi$ are Lagrange multipliers. They impose the chirality conditions $\p_-X^+=\p_+X^-=0$, so that in static gauge every solution has
\begin{equation}\label{eq:NCtheta}
t=\kappa\tau,\qquad \vartheta=\kappa\sigma+\sigma_0,\qquad \kappa\in\mathbb Z_{>0},
\end{equation}
where $\kappa$ is the number of times the string winds the meridian. The theory then reduces to a free transverse field,
\begin{equation}\label{eq:NClin}
\p_\tau\big(\sin^2\vartheta\,\dot\varphi\big)=\p_\sigma\big(\sin^2\vartheta\,\varphi'\big),
\end{equation}
and the Virasoro constraints fix the multipliers algebraically. The spacetime energy, which is an integral of the multiplier $\xi_1=\xi-\bar\xi$ (appendix~\ref{app:NC}), then becomes the worldsheet energy of $\varphi$ divided by $\kappa$,
\begin{equation}\label{eq:Eintr}
E=\frac{T}{2\kappa}\oint\sin^2\vartheta\,\big(\dot\varphi^2+\varphi'^2\big)\,d\sigma ,
\end{equation}
while the angular momentum is $J=T\oint\sin^2\vartheta\,\dot\varphi\,d\sigma$.

\section{Intrinsic NR reduction}\label{sec:intrinsic}

\subsection{The Pohlmeyer field}\label{phlf}

Let us first construct the reduction entirely within the intrinsic theory. In the relativistic case the Virasoro constraints fix the lengths of $\p_\pm\vec X$, and the Pohlmeyer field is the only invariant left in the pair of tangents. The sNC analogue is immediate: by \eqref{eq:NCtheta}, the longitudinal projections of the tangents are fixed,
\begin{equation}\label{eq:longproj}
\tau^0(\p_\pm X)=\frac\kappa2,\qquad \tau^1(\p_\pm X)=\pm\frac\kappa2,
\end{equation}
and the only remaining data are the transverse projections $u_\pm\equiv E^2(\p_\pm X)=\sin\vartheta\,\p_\pm\varphi$. The natural invariant is the oriented area spanned by the tangents in the plane of the spatial longitudinal and transverse directions, normalised by their longitudinal projections,\footnote{In the relativistic theory, with vielbeins $e^0=c\,dt$ and spatial metric $G_S=c^2d\theta^2+\sin^2\theta\,d\phi^2$, the Pohlmeyer angle is $\cos\psi=G_S(\p_+X,\p_-X)/\big(e^0(\p_+X)\,e^0(\p_-X)\big)$, which involves only the spatial metric. However there could be another invariant normalised area in the relativistic theory, see appendix \ref{app:chirel}.  In the intrinsic sNC theory the situation is different, because the metric is degenerate. The Virasoro constraints do not fix the lengths, only longitudinal projections.}
\begin{equation}\label{eq:chicov}
\chi\equiv\frac{(\tau^1\wedge E^2)(\p_+X,\p_-X)}{2\,\tau^0(\p_+X)\,\tau^0(\p_-X)}=\frac{\sin\vartheta\,\dot\varphi}{\kappa}.
\end{equation}
The field $\chi$ is invariant under the transverse isometry, and under conformal reparametrisations $\sigma^\pm\to f^\pm(\sigma^\pm)$, since numerator and denominator scale in the same way. In static gauge the residual reparametrisations are only constant shifts, so this invariance matters before the static gauge is imposed. It makes $\chi$ independent of the choice of conformal gauge, which is used later. The orientation of $\tau^1\wedge E^2$ fixes the sign of $\chi$, which matters for the identification of the Lax pairs. Physically, it is the transverse velocity of the string in units of the longitudinal frequency.  The other invariant, analogous area in the plane of $\tau^0$ and $E^2$ gives instead, up to sign, $\sin\vartheta\,\varphi'/\kappa$. As we now show, only \eqref{eq:chicov} obeys a closed local equation, and this singles it out as the Pohlmeyer field.

\subsection{Reduced equation and boundary conditions}

\begin{figure}[t]
\centering
\begin{tikzpicture}[>=stealth,scale=0.95]
  \begin{scope}
    \draw[thick] (0,0) circle (1.6);
    \draw[red!70!black,thin] plot[smooth] coordinates {(1.438,0.667) (1.430,0.714)};
    \draw[red!70!black,dashed,very thin] plot[smooth] coordinates {(1.430,0.714) (1.407,0.759) (1.368,0.804) (1.314,0.848) (1.245,0.889) (1.163,0.928) (1.069,0.964) (0.962,0.997) (0.845,1.027) (0.719,1.052) (0.585,1.073) (0.444,1.090) (0.299,1.102) (0.150,1.109) (0.000,1.111) (-0.150,1.109) (-0.299,1.102) (-0.444,1.090) (-0.585,1.073) (-0.719,1.052) (-0.845,1.027) (-0.962,0.997) (-1.069,0.964) (-1.163,0.928) (-1.245,0.889) (-1.314,0.848) (-1.368,0.804) (-1.407,0.759)};
    \draw[red!70!black,thin] plot[smooth] coordinates {(-1.407,0.759) (-1.430,0.714) (-1.438,0.667) (-1.430,0.621) (-1.407,0.575) (-1.368,0.530) (-1.314,0.486) (-1.245,0.445) (-1.163,0.406) (-1.069,0.370) (-0.962,0.337) (-0.845,0.308) (-0.719,0.282) (-0.585,0.261) (-0.444,0.244) (-0.299,0.232) (-0.150,0.225) (-0.000,0.223) (0.150,0.225) (0.299,0.232) (0.444,0.244) (0.585,0.261) (0.719,0.282) (0.845,0.308) (0.962,0.337) (1.069,0.370) (1.163,0.406) (1.245,0.445) (1.314,0.486) (1.368,0.530) (1.407,0.575) (1.430,0.621) (1.438,0.667)};
    \draw[very thick,blue!70!black] plot[smooth] coordinates {(0.000,1.522) (0.137,1.484) (0.272,1.429) (0.405,1.360) (0.533,1.275) (0.655,1.176) (0.770,1.064) (0.877,0.941) (0.974,0.807) (1.060,0.665) (1.135,0.515) (1.197,0.360) (1.246,0.201) (1.282,0.039) (1.303,-0.123) (1.311,-0.284) (1.303,-0.441) (1.282,-0.594) (1.246,-0.740) (1.197,-0.878) (1.135,-1.006) (1.060,-1.124) (0.974,-1.229) (0.877,-1.321) (0.770,-1.398) (0.655,-1.460)};
    \draw[blue!70!black,dashed,thin] plot[smooth] coordinates {(0.655,-1.460) (0.533,-1.505) (0.405,-1.535) (0.272,-1.547) (0.137,-1.543) (0.000,-1.522) (-0.137,-1.484) (-0.272,-1.429) (-0.405,-1.360) (-0.533,-1.275) (-0.655,-1.176) (-0.770,-1.064) (-0.877,-0.941) (-0.974,-0.807) (-1.060,-0.665) (-1.135,-0.515) (-1.197,-0.360) (-1.246,-0.201) (-1.282,-0.039) (-1.303,0.123) (-1.311,0.284) (-1.303,0.441) (-1.282,0.594) (-1.246,0.740) (-1.197,0.878) (-1.135,1.006) (-1.060,1.124) (-0.974,1.229) (-0.877,1.321) (-0.770,1.398) (-0.655,1.460)};
    \draw[very thick,blue!70!black] plot[smooth] coordinates {(-0.655,1.460) (-0.533,1.505) (-0.405,1.535) (-0.272,1.547) (-0.137,1.543) (0.000,1.522)};
    \fill (0,1.522) circle (1.3pt) node[above,font=\scriptsize] {$\vartheta=0$};
    \fill (0,-1.522) circle (1.3pt) node[below,font=\scriptsize] {$\vartheta=\pi$};
    \node[blue!70!black,font=\scriptsize,right] at (0.631,1.197) {$\vartheta$};
    \draw[very thick,green!50!black] plot[smooth] coordinates {(1.178,0.412) (1.105,0.383) (1.025,0.355) (0.937,0.330) (0.843,0.307) (0.743,0.287) (0.638,0.269) (0.529,0.254) (0.416,0.242) (0.300,0.232)};
    \fill[green!50!black] (0.300,0.232) circle (1.3pt);
    \node[green!50!black,font=\scriptsize,below=2pt] at (0.300,0.232) {$y$};
    \fill (1.178,0.412) circle (1.6pt);
    \node[font=\scriptsize,above left=-1pt] at (1.178,0.412) {$P$};
    \draw[->,thick,blue!70!black] (1.178,0.412) -- (1.447,-0.287) node[right,font=\scriptsize] {$\p_\vartheta$};
    \draw[->,thick,red!70!black] (1.178,0.412) -- (1.608,0.602) node[right,font=\scriptsize] {$\p_\varphi$};
    \node[font=\small] at (0,-2.25) {(a)};
  \end{scope}
  \begin{scope}[xshift=4.3cm]
    \draw[->,thick,red!70!black] (0,0) -- (1.5,0) node[right,font=\small] {$P_1$};
    \draw[->,thick,red!70!black] (0,0) -- (0,1.5) node[above,font=\small] {$P_2$};
    \fill (0,0) circle (1.3pt);
    \draw[->,thick,blue!70!black] (0.45,0) arc (0:270:0.45);
    \node[blue!70!black,font=\small] at (-0.62,0.42) {$J$};
    \node[font=\scriptsize,align=left,anchor=north west] at (0.25,-0.6) {$[J,P_1]=P_2$\\ $[J,P_2]=-P_1$\\ $[P_1,P_2]=0$};
    \node[font=\small] at (0.75,-2.25) {(b)};
  \end{scope}
\end{tikzpicture}
\caption{(a) The meridian (blue), along which the string is stretched in static gauge. Also visualises the latitude circle (red) through a point $P$ of it, with the longitudinal tangent $\p_\vartheta$, the transverse tangent $\p_\varphi$, and the transverse displacement $y=\sin\vartheta\,\varphi$ (green), measured along the latitude circle. The dashed lines lie on the far side of the sphere. (b) The algebra $e(2)$ of the moving frame \eqref{eq:frame}: the rotation $J$, whose component in the Lax pair \eqref{eq:LaxNC} is the longitudinal motion $\p_\pm\vartheta\,J$ at $\lambda=1$, and the commuting translations $P_{1,2}$, which carry the transverse data. The colours indicate this correspondence.}
\label{fig:frame}
\end{figure}
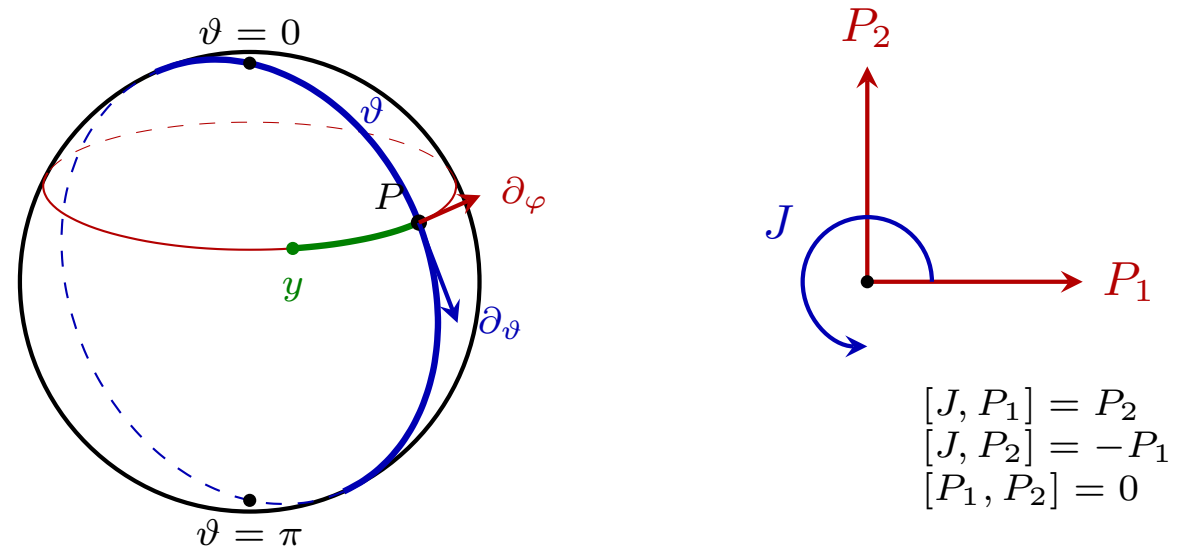

Let us introduce the transverse displacement $y=\sin\vartheta\,\varphi$, measured along the latitude circle (figure~\ref{fig:frame}a). With \eqref{eq:NCtheta}, a short computation turns \eqref{eq:NClin} into an equation with constant coefficients, $\ddot y-y''-\kappa^2y=0$. Since \eqref{eq:chicov} gives $\chi=\dot y/\kappa$, the Pohlmeyer field obeys
\begin{equation}\label{eq:chiLO}
\ddot\chi-\chi''-\kappa^2\chi=0,\qquad\text{i.e.}\qquad \p_+\p_-\chi=\frac{\kappa^2}{4}\,\chi ,
\end{equation}
which is a tachyonic Klein-Gordon equation. The gradient coming from the other invariant $\sin\vartheta\,\varphi'=y'-\kappa\cot\vartheta\,y$, by contrast, involves $\cot\vartheta$ explicitly and obeys no closed equation of this kind. The tachyonic mass reflects the tendency of the maximally stretched string to contract. It is the linearisation of the sine-Gordon equation about its unstable vacuum, see appendix~\ref{app:vacua}.

Finite energy requires $y$, and hence $\chi$, to vanish at the poles where $\sin\vartheta=0$, i.e. $\vartheta=j\pi$. This is since the energy \eqref{eq:Eintr} written in terms of $y$ contains boundary terms $\propto y^2\cot\vartheta$ at the poles (appendix~\ref{app:NC}). Equation \eqref{eq:chiLO} then needs complementary Dirichlet conditions, set by the longitudinal frequency. The modes are
\begin{equation}\label{eq:chimodes}
\begin{gathered}
\chi\propto\sin(\omega_n\tau+\beta_n)\,\sin n(\vartheta-j\pi),\\
\omega_n=\kappa\sqrt{n^2-1},\qquad n\ge1 .
\end{gathered}
\end{equation}
All frequencies here are real. Note that the string passes through $2\kappa$ poles as $\sigma$ runs around it, and at each of them $\chi$ is forced to vanish. The poles therefore act as fixed walls, at which Dirichlet conditions are imposed, and they cut the closed string into $2\kappa$ segments, the half-meridians between consecutive poles, each of which is a vibrating string with fixed ends. The Dirichlet walls on this segmented system thus remove the tachyonic instability, which survives only as the zero mode $n=1$, the rigid rotation.  They convert the unstable long-wavelength modes into configurations that are geometrically impossible on the sphere, because the transverse circle has zero size at the poles. The shortest mode that survives  sits exactly at the threshold, with zero frequency. The singular Sturm-Liouville structure and subtleties thereof will be discussed elsewhere \cite{Banerjee:Upcoming}.

To summarise in words, although the target-space geometry is non-relativistic, the worldsheet retains its two-dimensional Lorentz invariance (or conformal symmetry in the conformal gauge), and the reduced equation is a relativistic wave equation on the worldsheet. The tachyonic sign is a target-space effect: it reflects the instability of the maximally stretched meridian string to contract toward a point, but the Dirichlet boundary conditions at the poles (imposed by finite energy) stabilise them.

\subsection{Intrinsic Lax pair}

The reduced equation \eqref{eq:chiLO} is linear, and its general solution is a superposition of the modes \eqref{eq:chimodes}. A zero-curvature representation adds a structural statement to this, i.e. the equation can be written as the flatness of a connection valued in the algebra of a moving frame, with a spectral parameter.
The relevant algebra is suggested by the geometry of the sphere. The rotations of the sphere form $so(3)$. In sNC case, the sphere itself is not deformed; rather, the NR string stays close to a great circle, along which it moves by angles of order one, while its transverse displacements subtend angles of order $1/c$ (sections~\ref{sec:expansion} and \ref{sec:laxexp}). The rotation within the meridian plane survives, while the two rotations that mix the transverse direction with the meridian plane become commuting translations. The frame algebra therefore contracts to the Euclidean algebra $e(2)$ (figure~\ref{fig:frame}b),
\begin{equation}\label{eq:e2}
[J,P_1]=P_2, ~[J,P_2]=-P_1,~
[P_1,P_2]=0 .
\end{equation}
Note that $e(2)$ (or $iso(2)$) is used here as the structure group of a moving frame, and not as a symmetry of the sNC background, whose isometries do not include $J$. It is indeed a semidirect product of a $so(2)$ with the abelian ideal as above.

We can see the $e(2)$-valued connection
\begin{equation}\label{eq:LaxNC}
\begin{aligned}
A_+&=\frac{\lambda\kappa}{2}\,J+\p_+\chi\,P_2,\\
A_-&=-\frac{\kappa}{2\lambda}\big(J+\chi\,P_1\big),
\end{aligned}
\end{equation}
with spectral parameter $\lambda$, is flat precisely on solutions of \eqref{eq:chiLO}. The curvature is
\begin{equation}\label{eq:LaxNCcurv}
\p_-A_+-\p_+A_-+[A_+,A_-]=\Big(\p_+\p_-\chi-\frac{\kappa^2}{4}\chi\Big)P_2 .
\end{equation}
Flatness for all $\lambda$ is therefore equivalent to \eqref{eq:chiLO}. Note that the cancellation of the $P_1$ terms fixes the relative coefficient of $J$ and $\chi P_1$ in $A_-$, while the vanishing of the $P_2$ term is the equation of motion itself.

Let us now show that \eqref{eq:LaxNC} is not an ansatz, but the Maurer-Cartan form of an intrinsic moving frame. Consider the group element
\begin{equation}\label{eq:frame}
\begin{gathered}
\mathcal F=e^{\Upsilon}\,e^{\vartheta J},\\
\Upsilon\equiv-\frac y2\,P_1+\frac{\p_+y}{\kappa}\,P_2 ,
\end{gathered}
\end{equation}
which rotates by the longitudinal angle $\vartheta$ and translates by the transverse displacement $y=\sin\vartheta\,\varphi$ and by its light-cone derivative ("velocity"). Since $\Upsilon$ is a combination of commuting translations, we have $(\p_\pm e^{\Upsilon})e^{-\Upsilon}=\p_\pm \Upsilon$. Moreover, $e^{\Upsilon}Je^{-\Upsilon}=J+[\Upsilon,J]$ holds exactly, since $[\Upsilon,[\Upsilon,J]]$ is a commutator of translations and vanishes. By \eqref{eq:e2},
\begin{equation}\label{eq:XJ}
[\Upsilon,J]=\frac y2\,P_2+\frac{\p_+y}{\kappa}\,P_1 ,
\end{equation}
and the Maurer-Cartan form, before using any gauge, coming from \eqref{eq:frame} is
\begin{equation}\label{eq:MCgeneral}
\begin{aligned}
\p_\pm\mathcal F\,\mathcal F^{-1}=\;&\p_\pm \Upsilon+\p_\pm\vartheta\,J\\
&+\p_\pm\vartheta\Big(\frac y2\,P_2+\frac{\p_+y}{\kappa}\,P_1\Big).
\end{aligned}
\end{equation}

In static gauge, $\p_\pm\vartheta=\pm\frac\kappa2$. In the $\p_+$ component of \eqref{eq:MCgeneral}, the coefficient of $P_1$ vanishes identically, while the coefficient of $P_2$ simplifies on solutions of the transverse equation, $\p_+\p_-y=\frac{\kappa^2}{4}y$:
\begin{equation}\label{eq:MCplus}
\frac1\kappa\,\p_+^2y+\frac\kappa4\,y=\frac1\kappa\,\p_+\big(\p_++\p_-\big)y=\frac1\kappa\,\p_+\dot y=\p_+\chi .
\end{equation}
In the $\p_-$ component, the coefficient of $P_2$ vanishes by the same equation, and the coefficient of $P_1$ is
\begin{equation}\label{eq:MCminus}
-\frac12\big(\p_++\p_-\big)y=-\frac12\,\dot y=-\frac\kappa2\,\chi .
\end{equation}
Hence the Maurer-Cartan form is
\begin{equation}\label{eq:MCresult}
\begin{aligned}
\p_+\mathcal F\,\mathcal F^{-1}&=\frac\kappa2\,J+\p_+\chi\,P_2,\\
\p_-\mathcal F\,\mathcal F^{-1}&=-\frac\kappa2\big(J+\chi\,P_1\big),
\end{aligned}
\end{equation}
which is precisely \eqref{eq:LaxNC} at $\lambda=1$. Flatness at $\lambda=1$ is thus an identity of the Maurer-Cartan form of a well-defined group-valued field.\footnote{For a general element $\mathcal F=\exp(aP_1+bP_2)\exp(\vartheta J)$, with $a$ and $b$ functions of the transverse displacement and its derivatives, the requirement of reproducing \eqref{eq:LaxNC} fixes $(a,b)$ up to a solution of the homogeneous system, $a+ib\propto e^{i\vartheta}$. This ambiguity is the right multiplication of $\mathcal F$ by a constant group element, which leaves the Maurer-Cartan form invariant. Up to it, $\mathcal F$ is \eqref{eq:frame}.} The equation of motion is needed only to express the components through $\chi$.

The spectral parameter also has an intrinsic origin, in the boost invariance of the reduced equation. Let us consider the worldsheet boost
\begin{equation}\label{eq:boost}
f:\;(\sigma^+,\sigma^-)\mapsto\big(e^\rho\sigma^+,\,e^{-\rho}\sigma^-\big).
\end{equation}
The equation \eqref{eq:chiLO} is invariant under \eqref{eq:boost}, so that $\chi\circ f$ is a solution whenever $\chi$ is, although the static-gauge background \eqref{eq:NCtheta} is not. The logic is then as follows. Starting from a solution $\chi$, we boost it to obtain a new solution $\chi\circ f$. This new solution has its own moving frame \eqref{eq:frame}, whose Maurer-Cartan form is the connection \eqref{eq:MCresult} at $\lambda=1$, now built from $\chi\circ f$. We then pull this connection back to the original worldsheet coordinates. Since $\sigma^\pm$ are rescaled by $e^{\pm\rho}$, the components $A_\pm$ of the pulled-back one-form are multiplied by $e^{\pm\rho}$, and this rescaling is precisely what turns the coefficients $\pm\kappa/2$ of $J$ into $\lambda\kappa/2$ and $-\kappa/(2\lambda)$. Explicitly, pulling back the connection \eqref{eq:LaxNC} by $f$, we find
\begin{equation}\label{eq:boostpull}
\begin{gathered}
e^{\rho}\,A_+\big|_{\lambda,\chi}\circ f=A_+\big|_{1,\chi\circ f},\\
e^{-\rho}\,A_-\big|_{\lambda,\chi}\circ f=A_-\big|_{1,\chi\circ f},\qquad \lambda=e^{-\rho},
\end{gathered}
\end{equation}
as we have verified. The family \eqref{eq:LaxNC} is therefore the pull-back, by worldsheet boosts, of the Maurer-Cartan form \eqref{eq:MCresult} of the frame of the boosted solution, and the spectral parameter is the exponential of the boost rapidity. Since \eqref{eq:MCresult} is flat for every solution and the pull-back preserves flatness, this also explains why \eqref{eq:LaxNC} is flat for all $\lambda$. The mechanism is the same as in the relativistic Pohlmeyer reduction, where the spectral parameter of the sine-Gordon Lax pair reflects the Lorentz invariance of the reduced equation.

The structure of \eqref{eq:LaxNC} now has a direct reading. Its $J$-part is the longitudinal motion along the meridian, equal to $\p_\pm\vartheta\,J$ at $\lambda=1$, while its translational part carries the transverse dynamics through $\chi$. The $J$-part alone is flat, being constant, whereas the translational part alone is not. The flatness of the sum requires the cancellation in \eqref{eq:LaxNCcurv}, which is precisely the reduced equation. Note, however, that integrability is not by itself a strong statement for the linear equation \eqref{eq:chiLO}, since sNC sigma models that reduce to free fields in static gauge are trivially integrable \cite{FN2}. The content of \eqref{eq:LaxNC} is rather that the reduced theory carries a Lax pair valued in the contraction of the relativistic one.

Two caveats should be recorded. First, the frame \eqref{eq:frame} is not unique. The frame that matches $y$ and $y'$ at each point by a rigid rotation of the meridian yields a different pair, which is also flat precisely on \eqref{eq:chiLO}, but is not related to \eqref{eq:LaxNC} by a gauge transformation local in $y$. The pair \eqref{eq:LaxNC} is singled out by its coincidence with the LO and NLO sine-Gordon Lax pairs, which we will elaborate on in section~\ref{sec:expansion}.\footnote{The other frame in question is $\tilde{\mathcal F}=\exp(b_1P_1+b_2P_2)\exp(\vartheta J)$, with $(b_1,b_2)^T=R(\vartheta)\,(y,\,y'/\kappa)^T$ and $R(\vartheta)$ the rotation by $\vartheta$, so that the rigid rotation it describes has the same $y$ and $y'$ as the string at each point. On solutions of \eqref{eq:chiLO}, its Maurer-Cartan form, written in the conventions of \eqref{eq:LaxNC} and with the spectral parameter inserted by the boost \eqref{eq:boost}, is
$\tilde A_+=-\frac{\lambda\kappa}{2}\big(J+\chi P_1\big)-\p_+\chi\,P_2$ and
$\tilde A_-=\frac{\kappa}{2\lambda}\big(J-\chi P_1\big)+\p_-\chi\,P_2$,
with curvature $-2\big(\p_+\p_-\chi-\frac{\kappa^2}{4}\chi\big)P_2$. We have checked that no translation gauge transformation with parameters linear in $y$ and its first derivatives relates this pair to \eqref{eq:LaxNC}, even after composing with the automorphisms of $e(2)$ that reverse $J$ or the translations.} Second, we have not compared \eqref{eq:LaxNC} with the sigma-model Lax pairs of \cite{Kluson,RoyLax,Fontanella:2020eje,Fontanella:2022fjd}, which are constructed from the sNC sigma model directly rather than from the reduced field.

\subsection{Monodromy and conserved charges}\label{sec:monodromy}

Let us first recall why a monodromy produces conserved charges. A flat connection defines a parallel transport, and transporting a frame once around the closed string, from $\sigma=0$ to $\sigma=2\pi$, returns it in general transformed by a group element, the monodromy $M(\lambda)$. Flatness implies that changing the time at which the loop is taken only conjugates $M(\lambda)$, so that every quantity invariant under conjugation is conserved. In the relativistic $\mathfrak{su}(2)$ theory, such invariants are the eigenvalues of the monodromy matrix $M_{\rm rel}(\lambda)$ along $\sigma\in[0,2\pi)$, and expanding them in $\lambda$ generates the tower of conserved charges. In the sNC case, the flatness of the $\mathfrak{e}(2)$ Lax pair \eqref{eq:LaxNC} similarly allows us to construct conserved quantities from its spatial monodromy. With $\sigma^\pm=\tau\pm\sigma$, the spatial component of the connection is $A_\sigma=A_+-A_-$, and \eqref{eq:LaxNC} gives
\begin{equation}\label{eq:Asigma}
A_\sigma=a\,J+\frac{\kappa\chi}{2\lambda}\,P_1+\p_+\chi\,P_2,\qquad
a\equiv\frac{\kappa}{2}\Big(\lambda+\frac1\lambda\Big).
\end{equation}
Let $U(\sigma;\lambda)$ be the Wilson line along the string, $\p_\sigma U=A_\sigma U$ with $U(0;\lambda)=\mathbf 1$, and let $M(\lambda)\equiv U(2\pi;\lambda)$ be the monodromy. Factorising $U(\sigma;\lambda)=e^{a\sigma J}W(\sigma;\lambda)$, where $a$ is constant and
\begin{equation}
\p_\sigma W=e^{-a\sigma J}\big(\frac{\kappa\chi}{2\lambda}P_1+\p_+\chi\,P_2\big)e^{a\sigma J}W
\end{equation}
 takes values in the abelian ideal spanned by $P_1$ and $P_2$, the path-ordered exponential reduces to an ordinary integral:
\begin{equation}\label{eq:monodromy}
M(\lambda)=e^{2\pi aJ}\,e^{\Omega_1P_1+\Omega_2P_2},
\end{equation}
where, using $e^{-a\sigma J}(P_1-iP_2)e^{a\sigma J}=e^{-ia\sigma}(P_1-iP_2)$,
\begin{equation}\label{eq:Omega}
\Omega_1(\lambda)+i\,\Omega_2(\lambda)=\int_0^{2\pi}\Big(\frac{\kappa\chi}{2\lambda}+i\,\p_+\chi\Big)e^{-ia\sigma}\,d\sigma .
\end{equation}

The structure of \eqref{eq:monodromy} has a simple geometric reading in the transverse plane. The factor $e^{2\pi aJ}$ is a rotation by the angle $2\pi a$, produced by the longitudinal motion of the frame along the meridian, while $e^{\Omega_1P_1+\Omega_2P_2}$ is a translation by the vector $(\Omega_1,\Omega_2)$, produced by the transverse dynamics of the string. In the Euclidean plane, a rotation by an angle that is not a multiple of $2\pi$, followed by a translation, is again a rotation by the same angle, but about a shifted centre ( see figure~\ref{fig:monodromy}a). The translation can then be removed by moving the origin, and the monodromy remembers nothing of $\chi$. Only when the rotation is trivial, $2\pi a\in2\pi\mathbb Z$, does the translation survive as a genuine invariant, namely its length. We call the values of $\lambda$ at which this happens \textit{resonant}. Note also the meaning of the Fourier factor $e^{-ia\sigma}$ in \eqref{eq:Omega}: as the frame is transported along the string, it rotates at the rate $a$ per unit $\sigma$, and $\Omega_1+i\Omega_2$ collects the transverse data as seen in this rotating frame. A transverse wave whose wavenumber matches the rotation rate $a$ adds up coherently along the string, while all other waves average to zero. A resonance is thus a matching between the rotation rate of the frame, which depends on $\lambda$, and the wavenumber of a mode of the string. Since the smooth closed-string solutions contain only the wavenumbers $n\kappa$, the non-trivial resonances are those with $a=n\kappa$. We now make this precise.

For solutions that are smooth across the poles, flatness implies that the monodromy evolves by conjugation, $M(\tau)=g(\tau)M(0)g(\tau)^{-1}$, with $g$ in the group $E(2)$ generated by \eqref{eq:e2}. For generic $\lambda$, as mentioned before, $e^{2\pi aJ}\ne\mathbf 1$ and $M(\lambda)$ is conjugate to the pure rotation $e^{2\pi aJ}$, whose conjugacy class carries no information about $\chi$. At the resonant values of $\lambda$ where $a\in\mathbb Z$, however, $e^{2\pi aJ}=\mathbf 1$ and $M(\lambda)$ reduces to a pure translation. Conjugation by translations then leaves $(\Omega_1,\Omega_2)$ invariant, while conjugation by rotations only rotates the vector $(\Omega_1,\Omega_2)$ in the $(P_1,P_2)$ plane. Hence the squared norm
\begin{equation}\label{eq:monocharge}
\mathcal Q(\lambda)\equiv\Omega_1(\lambda)^2+\Omega_2(\lambda)^2,\qquad a\in\mathbb Z,
\end{equation}
is gauge invariant and conserved in $\tau$. 

Note that $M(\lambda)$ itself is always well defined, since $y$, and hence our field $\chi$, is continuous and vanishes at the poles. This makes $A_\sigma$ at worst piecewise continuous. Its conservation, however, requires flatness at the poles as well: if the segment modes \eqref{eq:chimodes} are excited independently on each half-meridian, $\chi'$ jumps at the poles. Consequently, the curvature acquires delta-function support there, and the poles act as sources in the evolution of $M(\lambda)$. Here we simplify and restrict to globally smooth periodic solutions, which are formed by segment modes $\sin n(\kappa\sigma+\sigma_0)$ with alternating signs across consecutive half-meridians.

To evaluate the resonant charges \eqref{eq:monocharge} on such global solutions, consider a mode of wavenumber $p=n\kappa$ ($n\in\mathbb Z_{>0}$), decomposed into right- and left-moving components:
\begin{equation}\label{eq:modeLR}
\begin{gathered}
\chi(\sigma,\tau)= A_n\,e^{i(n\kappa\sigma-\omega_n\tau)}+B_n\,e^{i(n\kappa\sigma+\omega_n\tau)}+\mathrm{c.c.},\\
\omega_n=\kappa\sqrt{n^2-1}.
\end{gathered}
\end{equation}
In the Fourier integral \eqref{eq:Omega}, the factor $e^{-ia\sigma}$ selects the $e^{in\kappa\sigma}$ harmonic when $a=n\kappa\in\mathbb Z_{>0}$, i.e.\ $\lambda+1/\lambda=2n$, whose two roots are
\begin{equation}\label{eq:resonance}
\lambda_n^\pm=n\pm\sqrt{n^2-1}=e^{\pm\theta_n},\qquad \cosh\theta_n=n ,
\end{equation}
while the integral over $\sigma\in[0,2\pi)$ annihilates the conjugate $e^{-in\kappa\sigma}$ terms (as well as all modes with $m\ne n$). Acting with $\frac{\kappa}{2\lambda}+i\p_+$ on \eqref{eq:modeLR}, where $i\p_+=\frac i2(\p_\tau+\p_\sigma)$, multiplies $A_n e^{i(n\kappa\sigma-\omega_n\tau)}$ by $\frac{\kappa}{2\lambda}+\frac12(\omega_n-n\kappa)$ and $B_n e^{i(n\kappa\sigma+\omega_n\tau)}$ by $\frac{\kappa}{2\lambda}-\frac12(\omega_n+n\kappa)$. Using $\omega_n=\kappa\sqrt{n^2-1}$, the roots \eqref{eq:resonance} satisfy
\begin{equation}
n\kappa-\omega_n=\kappa\lambda_n^-=\frac{\kappa}{\lambda_n^+},\qquad
n\kappa+\omega_n=\kappa\lambda_n^+=\frac{\kappa}{\lambda_n^-}.
\end{equation}
Substituting these into \eqref{eq:Omega} at $a=n\kappa$ and integrating over $\sigma\in[0,2\pi)$ gives
\begin{equation}
\begin{aligned}
\Omega_1(\lambda)+i\Omega_2(\lambda)\Big|_{a=n\kappa}
&=\pi\kappa\left(\frac{1}{\lambda}-\frac{1}{\lambda_n^+}\right)A_n e^{-i\omega_n\tau}\\
&\quad+\pi\kappa\left(\frac{1}{\lambda}-\frac{1}{\lambda_n^-}\right)B_n e^{i\omega_n\tau}.
\end{aligned}
\end{equation}
Evaluating at the two roots \eqref{eq:resonance} therefore projects onto each chirality separately: $\lambda=\lambda_n^+$ annihilates the right-mover $A_n$, while $\lambda=\lambda_n^-$ annihilates the left-mover $B_n$. Using $\frac{1}{\lambda_n^-}-\frac{1}{\lambda_n^+}=\lambda_n^+-\lambda_n^-=2\omega_n/\kappa$, we find
\begin{equation}
\begin{aligned}
\Omega_1(\lambda_n^-)+i\Omega_2(\lambda_n^-)&=2\pi\omega_n\,A_n e^{-i\omega_n\tau},\\
\Omega_1(\lambda_n^+)+i\Omega_2(\lambda_n^+)&=-2\pi\omega_n\,B_n e^{i\omega_n\tau}.
\end{aligned}
\end{equation}
Taking the squared modulus \eqref{eq:monocharge} removes the time-dependent phases $e^{\mp i\omega_n\tau}$ and yields
\begin{equation}\label{eq:monochargeAB}
\mathcal Q(\lambda_n^-)=4\pi^2\omega_n^2\,|A_n|^2,\qquad \mathcal Q(\lambda_n^+)=4\pi^2\omega_n^2\,|B_n|^2 .
\end{equation}
The two resonant roots \eqref{eq:resonance} thus separate the two directions of propagation and reproduce the quadratic oscillator invariants of the linear reduced theory, one for each mode and chirality.\footnote{Note in this section that having only $e(2)$ Lax representation somehow sidesteps the problem found for full sNC strings on AdS$\times$S \cite{Fontanella:2022wfj,Fontanella:2026gaq}, where the monodromy eigenvalues turn out to be independent of the spectral parameter.}

For $n=1$, the two roots coalesce at $\lambda_1^\pm=1$ where $\omega_1=0$, so that $\mathcal Q(1)=0$. The static $n=1$ mode of \eqref{eq:chimodes}, i.e. the rigid rotation of the meridian, is therefore invisible to the $\mathfrak{e}(2)$ monodromy, which does not capture the spacetime angular momentum $J$. Geometrically, at $\lambda=1$ the Lax connection is the Maurer-Cartan form of the frame \eqref{eq:frame}, so that $U(\sigma;1)=\mathcal F(\sigma)^{-1}\mathcal F(0)$, and for solutions smooth across the poles the frame is single-valued around the closed string, giving $M(1)=\mathbf 1$. The geometric and kinematic meaning of the resonant charges is discussed further in appendix~\ref{sec:monophys}.\subsection{Explicit maps}

\begin{figure*}[t]
\centering
\includegraphics[width=0.85\textwidth]{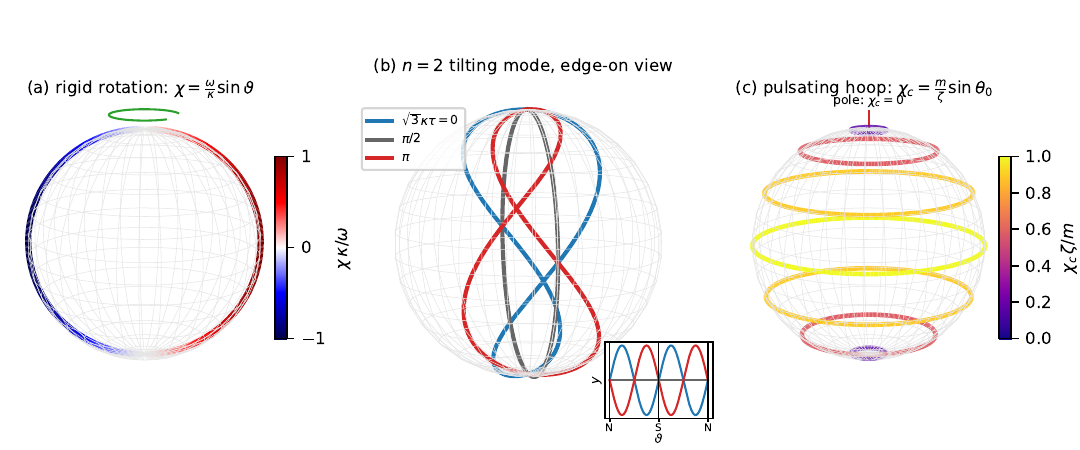}
\caption{NR strings and their Pohlmeyer fields on the unit sphere. (a) The rigidly rotating string coloured by $\chi$, which vanishes at the poles and reaches $\pm\omega/\kappa$ on the equator. (b) The tilting oscillator \eqref{eq:chin2} at three instants, viewed almost edge-on to the meridian plane so that the transverse (azimuthal) displacement appears sideways; it is exaggerated for visibility ($a=0.8$). The inset shows the same mode in the sNC variables, transverse displacement $y=\sin\vartheta\,\varphi$ against the longitudinal angle $\vartheta$: a standing wave between the poles. (c) The pulsating string of the momentum sector at successive times, coloured by $\chi_c$, which vanishes as the hoop collapses through a pole.}
\label{fig:strings}
\end{figure*}

In terms of our reduced fields $y$ and $\chi$, the energy \eqref{eq:Eintr} reads
\begin{equation}\label{eq:Ewind}
E=\frac{T\kappa}{2}\oint\Big[\chi^2+\big(\p_\vartheta y\big)^2-y^2\Big]d\sigma .
\end{equation}
The kinetic part is quadratic in the Pohlmeyer field, while the remaining terms are the energy stored in the spatial profile of the tachyonic field. On top of the static gauge, choosing a special classical profile, for the rigidly rotating string, i.e. $\varphi=\omega\tau$, \eqref{eq:chicov} gives
\begin{equation}\label{eq:chirot}
\chi=\frac\omega\kappa\sin\vartheta,\qquad E=\frac{T\kappa}{2}\oint\chi^2\,d\sigma=\frac{J^2}{2\pi\kappa T},
\end{equation}
since the gradient terms integrate to zero. This is the static mode $n=1$ of \eqref{eq:chimodes}, and it reproduces the Galilean ``rigid rotor" dispersion relation (figure~\ref{fig:strings}a), with $J=\pi T\omega$. The true first oscillator mode, $n=2$, is
\begin{equation}\label{eq:chin2}
\begin{aligned}
\chi&=-\frac{\sqrt3\,a}{2}\sin\big(\sqrt3\kappa\tau\big)\sin2\vartheta,\\
\varphi&=\varphi_0+a\cos\big(\sqrt3\kappa\tau\big)\cos\vartheta ,
\end{aligned}
\end{equation}
a transverse tilting oscillation, i.e. a standing wave between the poles, of the meridian with energy $\frac{3\pi}{8}T\kappa a^2$. Its shape is best understood in terms of the transverse displacement, $y=\frac a2\cos(\sqrt3\kappa\tau)\sin2\vartheta$. Each point of the meridian is rotated about the polar axis by an angle proportional to $\cos\vartheta$, but the displacement vanishes at the poles and at the equator. The two halves of each segment are therefore pushed in opposite transverse directions, and the meridian bends periodically into an S-shape (figure~\ref{fig:strings}b).

Conversely, given $\chi$, \eqref{eq:chicov} gives $\dot\varphi=\kappa\chi/\sin\vartheta$, and \eqref{eq:NClin} then yields
\begin{equation}\label{eq:invert}
\sin^2\vartheta\,\varphi'=\kappa\int\sin\vartheta\,\dot\chi\,d\sigma+v ,
\end{equation}
whose compatibility condition is equivalent to the reduced equation \eqref{eq:chiLO}. Every solution of the reduced equation therefore lifts to a string, up to the transverse isometry and the constant $v$. The latter is the twist charge of the twisted rotating strings $\varphi=\omega\tau+f(\sigma)$, with $\sin^2\vartheta\,f'=v$, which have the same $\chi$ as the untwisted string.\footnote{For $\varphi=\omega\tau+f(\sigma)$, the reduced field is
$\chi=\sin\vartheta\,\dot\varphi/\kappa=(\omega/\kappa)\sin\vartheta$,
which is independent of $f$: the twist is invisible to $\chi$. The
transverse equation \eqref{eq:NClin} then becomes
$\partial_\sigma(\sin^2\vartheta\,f')=0$, so that
$\sin^2\vartheta\,f'=v$ is a constant. This is exactly the integration
constant of the inverse map \eqref{eq:invert}. The constant $v$ is thus
a conserved charge of the transverse equation, the twist charge, and
it parametrises the many-to-one nature of the reduction.} The reduction is thus many-to-one, and the singularity of the twisted strings at the poles, where $f'\simeq v/(\vartheta-\vartheta_*)^2$ (appendix~\ref{app:NC}), resides entirely in the inverse map.

\section{The $1/c^2$ expansion}\label{sec:expansion}

\subsection{The setup}

Let us now turn to the expansion around $c\to\infty$, which gives a systematic handle over the solution space \cite{Relating,HH1,HH2,Hartong:2024ydv}. Note that this $c$ acts as speed of light in the target space, and we consider our fields analytic in a $c$ expansion. Once expanded, the solution space is actually larger than that of the intrinsic theory (appendix~\ref{app:larger}). In this section and the next we also follow relativistic solutions to their NR counterparts, in the spirit of \cite{Fontanella:2023men}, where string vacua were tracked from relativistic to non-relativistic $AdS_5\times S^5$. The expansion is performed on the background
\begin{equation}\label{eq:metc}
\begin{aligned}
ds^2&=c^2\big(-dt^2+d\theta^2\big)+\sin^2\theta\,d\phi^2\\
&=c^2\Big[-dt^2+d\theta^2+\sin^2\theta\,d\phi_{\rm rel}^2\Big],
\end{aligned}
\end{equation}
with $\phi_{\rm rel}=\phi/c$. Locally this is $\mathbb R\times S^2$ with a sphere of radius $c$, so the reduction of section~\ref{sec:pohl} applies exactly at finite $c$, with $X^0=ct$.

We should mention one subtlety here. Globally there are two completions. If $\phi$ has period $2\pi$, the sphere is squashed, with $U(1)$ isometry, and its poles are conical singularities of deficit angle $2\pi(1-1/c)$. If instead $\phi_{\rm rel}$ has period $2\pi$, the sphere is round, with $so(3)$ isometry, and $\phi$ has period $2\pi c$. The Pohlmeyer reduction is local and holds in both. We should physically think of this in a way that on this sphere meridian strings move at a speed of $O(1)$ while the transverse speed is of $O(1/c)$. This asymmetry makes the leading theory a zoomed in version around the neighbourhood of the static great circle\footnote{One can compare this with the pp-wave limit, for example.}.  The comparison with the elliptic solutions of \cite{KMP1} below uses the round completion, while the non-relativistic analysis of sections~\ref{sec:intrinsic} and the current one is insensitive to the choice, except for the closure conditions noted below.

Three kinds of objects are expanded. First, the worldsheet fields and the worldsheet metric are expanded in even powers of $1/c$ \cite{HH1,HH2},
\begin{equation}\label{eq:fieldexp}
t=t_0+\frac{t_1}{c^2},\quad \theta=\theta_0+\frac{\theta_1}{c^2},\quad \phi=\phi_0+\frac{\phi_1}{c^2},
\end{equation}
upto $O(c^{-4})$ terms, so that the Polyakov Lagrangian becomes
\begin{equation}\label{eq:Lexp}
\begin{gathered}
\mathcal L=c^2\,\mathcal L_{\rm LO}+\mathcal L_{\rm NLO}+O(c^{-2}),\\
\mathcal L_{\rm LO}=\frac T2\big(t_0'^2-\dot t_0^2-\theta_0'^2+\dot\theta_0^2\big),
\end{gathered}
\end{equation}
with $\mathcal L_{\rm NLO}$ given in appendix~\ref{A3}. The LO Lagrangian contains only the longitudinal fields $t_0$ and $\theta_0$. The transverse field $\phi_0$ first appears in $\mathcal L_{\rm NLO}$, together with the subleading longitudinal fields $t_1$ and $\theta_1$, which play the role of the intrinsic multipliers. The field $\phi_1$, and corrections to $t_1$ and $\theta_1$, enter only at NNLO. Second, the Pohlmeyer field $\psi$ is expanded, and we shall see in section~\ref{sec:windexp} that it runs in odd powers of $1/c$. Third, the Lax pair \eqref{eq:Lax} and its curvature are expanded, which we do in section~\ref{sec:laxexp}.

At LO the string sees only the longitudinal cylinder $(t,\theta)$, and the LO Virasoro constraint is $\p_\pm X_0^+\,\p_\pm X_0^-=0$, with $X_0^\pm=t_0\pm\theta_0$. It is a product of two light-cone derivatives, which can vanish in two inequivalent ways. In the first, $\p_-X_0^+=\p_+X_0^-=0$, so that $X_0^\pm$ are chiral and, in static gauge, $t_0=\kappa\tau$ and $\theta_0=\kappa\sigma+\sigma_0$: the string is stretched along the meridian and winds it $\kappa\in\mathbb Z_{>0}$ times, and its pulled-back longitudinal metric is non-degenerate. We call this the winding sector. It is, as we shall see, the sector described by the intrinsic theory, with $\kappa$ the winding of \eqref{eq:NCtheta}, and on it the intrinsic variables map: $\theta_0=\vartheta$ and $\phi_0=\varphi$.

In the second, $X_0^-$ is constant, so that $t_0=\theta_0=\zeta\tau+\alpha_0$: the string has no winding, is point-like in the longitudinal plane, and moves along the longitudinal circle at the speed of light, with a frequency $\zeta$. We call this the momentum sector. Its pulled-back longitudinal metric vanishes, and it has no intrinsic sNC description. The third possibility, $\p_+X_0^+=\p_-X_0^-=0$, is the image of the winding sector under worldsheet parity, $\sigma\to-\sigma$, which simply reverses the orientation of the string, and it adds no new configurations (appendix~\ref{A3}).

The LO dynamics is that of a string in the two-dimensional longitudinal target, and it has no local degrees of freedom: it selects a branch and fixes the longitudinal embedding up to global data. The two branches sit at the two vacua of the sine-Gordon equation (appendix~\ref{app:vacua}): in the winding sector $|\p_\sigma\vec X|$ is maximal and $\psi\to\pi$, the stretched string at rest, while in the momentum sector $|\p_\tau\vec X|$ is maximal and $\psi\to0$, the point-like string moving at the speed of light. The LO branch therefore decides about which vacuum the expansion is performed, and the Pohlmeyer field and its dynamics first appear at NLO.

In static gauge we write $t=\kappa_r\tau$ in the winding sector and $t=\zeta_r\tau$ in the momentum sector, with
\begin{equation}\label{eq:freqexp}
\kappa_r=\kappa+\frac{h}{c^2}+O(c^{-4}),\qquad \zeta_r=\zeta+O(c^{-2}).
\end{equation}
The quantities on the left-hand side are not new parameters in the parent theory. In the relativistic theory $t$ is a free field, so the residual conformal symmetry brings every closed string to the static gauge $t=\mu\tau$ with constant $\mu$, the mass of the sine-Gordon equation. The parameters $\kappa_r$ and $\zeta_r$ are the values of $\mu$ on solutions near the two vacua, whose LO values are the winding $\kappa$ and the light-cone frequency $\zeta$ respectively.\footnote{On a closed string $t$ is $\mu\tau$ plus periodic left- and right-moving pieces, which the residual conformal reparametrisations remove. In the expansion these reparametrisations are of order $c^{-2}$: they leave $t_0=\kappa\tau$ untouched and set $t_1=h\tau$, with $h=E_{\rm NLO}/(2\pi T)$ fixed by the NLO Virasoro constraint (appendix~\ref{app:levelmatch}). For the rigid rotation, $h=\omega^2/(4\kappa)$.}

In static gauge, the identification with the relativistic calculations on the sphere in section~\ref{sec:pohl} is
\begin{equation}\label{eq:ident}
R=c,\qquad \mu=\kappa_r ,
\end{equation}
with $\mu=\zeta_r$ in the momentum sector.
The sine-Gordon mass is thus the longitudinal frequency, independent of $c$. The relations \eqref{eq:static} on the relativistic case become
\begin{equation}\label{eq:staticc}
\begin{aligned}
c^2\dot\theta^2+\sin^2\theta\,\dot\phi^2&=c^2\kappa_r^2\cos^2\frac\psi2,\\
c^2\theta'^2+\sin^2\theta\,\phi'^2&=c^2\kappa_r^2\sin^2\frac\psi2 ,
\end{aligned}
\end{equation}
and these exact relations locate the expansion on the sine-Gordon side. Note that the non-relativistic analogue of the BMN particle, point-like and moving at the speed of light, lies on the momentum branch. By contrast, the BMN-like strings of non-relativistic $AdS_5\times S^5$ \cite{FN2} are extended along the compact longitudinal direction, and they are analogues of the winding sector.

\subsection{Winding sector: the Pohlmeyer field}\label{sec:windexp}
Expansion of the Pohlmeyer field is subtle compared to other quantities.
On the winding sector, in the NR variables, $\theta'=\kappa+O(c^{-2})$ and $\dot\theta=\dot\theta_1/c^2$. The second relation in \eqref{eq:staticc} gives $\psi\to\pi$, while the first shows that the deviation is of order $1/c$,
\begin{equation}\label{eq:chiexp}
\frac c2\,(\pi-\psi)=\frac{\sin\theta_0\,\dot\phi_0}{\kappa}+O(c^{-2})=\chi+O(c^{-2}).
\end{equation}
The leading term of $\frac c2(\pi-\psi)$ is therefore exactly the intrinsic field at $c\to \infty$ \eqref{eq:chicov} provided  $\theta_0=\vartheta$ and $\phi_0=\varphi$. Its corrections, of relative order $c^{-2}$, involve the NLO longitudinal fields $\theta_1$, $t_1$ and the frequency shift $h$, together with the NNLO field $\phi_1$. Let us therefore organise the expanded Pohlmeyer field by orders,
\begin{equation}\label{eq:psiorders}
\psi=\pi-\frac2c\Big(\underbrace{\chi_{\rm NLO}}_{\rm NLO}+\frac{1}{c^2}\,\underbrace{\chi_{\rm NNLO}}_{\rm NNLO}\Big)+O(c^{-5}),
\end{equation}
where $\pi$ is the constant value of $\psi$ on the frozen longitudinal configuration, which carries no dynamics. On top of that, $\chi_{\rm NLO}\equiv\chi$, our reduced field, and $\chi_{\rm NNLO}$ contains the corrections just listed.

\begin{figure}[t]
\centering
\resizebox{\columnwidth}{!}{%
\begin{tikzpicture}[>={Stealth[length=1.8mm]},font=\scriptsize]

  \fill[blue!5,rounded corners=2pt]   (1.85,-4.55) rectangle (3.85,0.45);
  \fill[red!5,rounded corners=2pt]    (3.95,-4.55) rectangle (6.15,0.45);
  \fill[gray!8,rounded corners=2pt]   (6.25,-4.55) rectangle (8.55,0.45);

  \node[font=\scriptsize\bfseries,blue!70!black,align=center] at (2.85,0.15)
    {LO\\[-1pt]{\tiny frozen longitudinal}};
  \node[font=\scriptsize\bfseries,red!70!black,align=center]  at (5.05,0.15)
    {NLO\\[-1pt]{\tiny intrinsic sNC}};
  \node[font=\scriptsize\bfseries,black!75,align=center]      at (7.40,0.15)
    {NNLO\\[-1pt]{\tiny non-linear / $su(2)$}};

  \draw[gray!50,thin] (0.0,-0.15) -- (8.55,-0.15);

  \node[anchor=west,font=\scriptsize\bfseries,align=left] at (0.0,-0.55)
    {Action $\mathcal L$\\[-2pt]{\tiny\color{gray!80!black}(even in $1/c$)}};
  \node[blue!70!black] at (2.85,-0.55) {$c^2\mathcal L_{\rm LO}$};
  \draw[->,gray!70] (3.55,-0.55) -- (4.25,-0.55) node[midway,above=-1pt,font=\tiny] {$c^{-2}$};
  \node[red!70!black]  at (5.05,-0.55) {$c^0\mathcal L_{\rm NLO}$};
  \draw[->,gray!70] (5.80,-0.55) -- (6.50,-0.55) node[midway,above=-1pt,font=\tiny] {$c^{-2}$};
  \node[black!80]      at (7.40,-0.55) {$c^{-2}\mathcal L_{\rm NNLO}$};

  \draw[gray!30,very thin] (0.0,-0.95) -- (8.55,-0.95);

  \node[anchor=west,font=\scriptsize\bfseries,align=left] at (0.0,-1.45)
    {String fields\\[-2pt]{\tiny\color{gray!80!black}(even in $1/c$)}};
  \node[blue!70!black,align=center] at (2.85,-1.45)
    {$t_0,\,\theta_0$\\[-1pt]{\tiny $O(1)$}};
  \node[red!70!black,align=center]  at (5.05,-1.45)
    {$\phi_0\;[O(1)]$\\[-1pt]{\tiny $c^{-2}(t_1,\theta_1)\sim\xi$}};
  \node[black!80,align=center]      at (7.40,-1.45)
    {$c^{-2}\phi_1$\\[-1pt]{\tiny $c^{-4}(t_2,\theta_2)$}};

  \draw[->,red!70!black,thin] (5.05,-1.82) -- (5.05,-2.12);

  \draw[gray!30,very thin] (0.0,-1.95) -- (8.55,-1.95);

  \node[anchor=west,font=\scriptsize\bfseries,align=left] at (0.0,-2.45)
    {Pohlmeyer $\psi$\\[-2pt]{\tiny\color{red!70!black}(odd in $1/c$)}};
  \node[blue!70!black,align=center] at (2.85,-2.45)
    {$\psi_0=\pi$\\[-1pt]{\tiny $O(1)$ vacuum}};
  \draw[->,red!70!black] (3.65,-2.45) -- (4.15,-2.45) node[midway,above=-1pt,font=\tiny] {$c^{-1}$};
  \node[red!70!black,align=center]  at (5.05,-2.45)
    {$-\dfrac{2}{c}\chi_{\rm NLO}$\\[-1pt]{\tiny $O(c^{-1})$}};
  \draw[->,red!70!black] (5.90,-2.45) -- (6.45,-2.45) node[midway,above=-1pt,font=\tiny] {$c^{-2}$};
  \node[black!80,align=center]      at (7.40,-2.45)
    {$-\dfrac{2}{c^3}\chi_{\rm NNLO}$\\[-1pt]{\tiny $O(c^{-3})$}};

  \draw[gray!30,very thin] (0.0,-2.95) -- (8.55,-2.95);

  \node[anchor=west,font=\scriptsize\bfseries,align=left] at (0.0,-3.45)
    {Lax pair $A_\pm$\\[-2pt]{\tiny\color{gray!80!black}(even $\sigma_1$, odd $\sigma_{2,3}$)}};
  \node[blue!70!black,align=center] at (2.85,-3.45)
    {$A_\pm^{(0)}\propto\sigma_1$\\[-1pt]{\tiny $O(1)\sim J$}};
  \node[red!70!black,align=center]  at (5.05,-3.45)
    {$c^{-1}A_\pm^{(1)}\propto\sigma_{2,3}$\\[-1pt]{\tiny $O(c^{-1})\sim P_{1,2}$}};
  \node[black!80,align=center]      at (7.40,-3.45)
    {$c^{-2}A_\pm^{(2)}\;[\sigma_1]$\\[-1pt]{\tiny $c^{-3}A_\pm^{(3)}\;[\sigma_{2,3}]$}};

  \draw[green!50!black,thick,dashed,rounded corners=2pt]
    (1.95,-3.82) rectangle (6.05,-3.05);
  \node[green!50!black,fill=white,inner sep=1pt,font=\tiny\bfseries]
    at (4.0,-3.05) { $e(2)$ Lax pair \eqref{eq:LaxNC}};

  \draw[gray!30,very thin] (0.0,-3.92) -- (8.55,-3.92);

  \node[anchor=west,font=\scriptsize\bfseries,align=left] at (0.0,-4.25)
    {Curvature $F$\\[-2pt]{\tiny\color{red!70!black}($F^{(2k)}\equiv 0$)}};
  \node[blue!70!black] at (2.85,-4.25) {$F^{(0)}=0$};
  \node[red!70!black,align=center]  at (5.05,-4.25)
    {$c^{-1}F^{(1)}=0$\\[-2pt]{\tiny Eq.~\eqref{eq:chiLO}}};
  \node[black!80,align=center]      at (7.40,-4.25)
    {$c^{-3}F^{(3)}=0$\\[-2pt]{\tiny Eq.~\eqref{eq:chiNNLOeq}}};

\end{tikzpicture}%
}
\caption{Hierarchy of orders in the winding sector. The Polyakov action and the string fields expand in even powers of $1/c$, whereas the deviation $\pi-\psi$ of the Pohlmeyer angle and the non-vanishing curvature components $F^{(2k+1)}$ expand in odd powers of $1/c$. Because the transverse metric has no factor of $c^2$ in \eqref{eq:metc}, $\phi_0$ enters at $\mathcal L_{\rm NLO}$ alongside $c^{-2}(t_1,\theta_1)$ and builds $\chi_{\rm NLO}$ at $O(c^{-1})$ in $\psi$. The intrinsic $e(2)$ Lax pair \eqref{eq:LaxNC} combines the $O(1)$ longitudinal connection $A_\pm^{(0)}$ with the $O(c^{-1})$ transverse connection $c^{-1}A_\pm^{(1)}$.}
\label{fig:orders}
\end{figure}
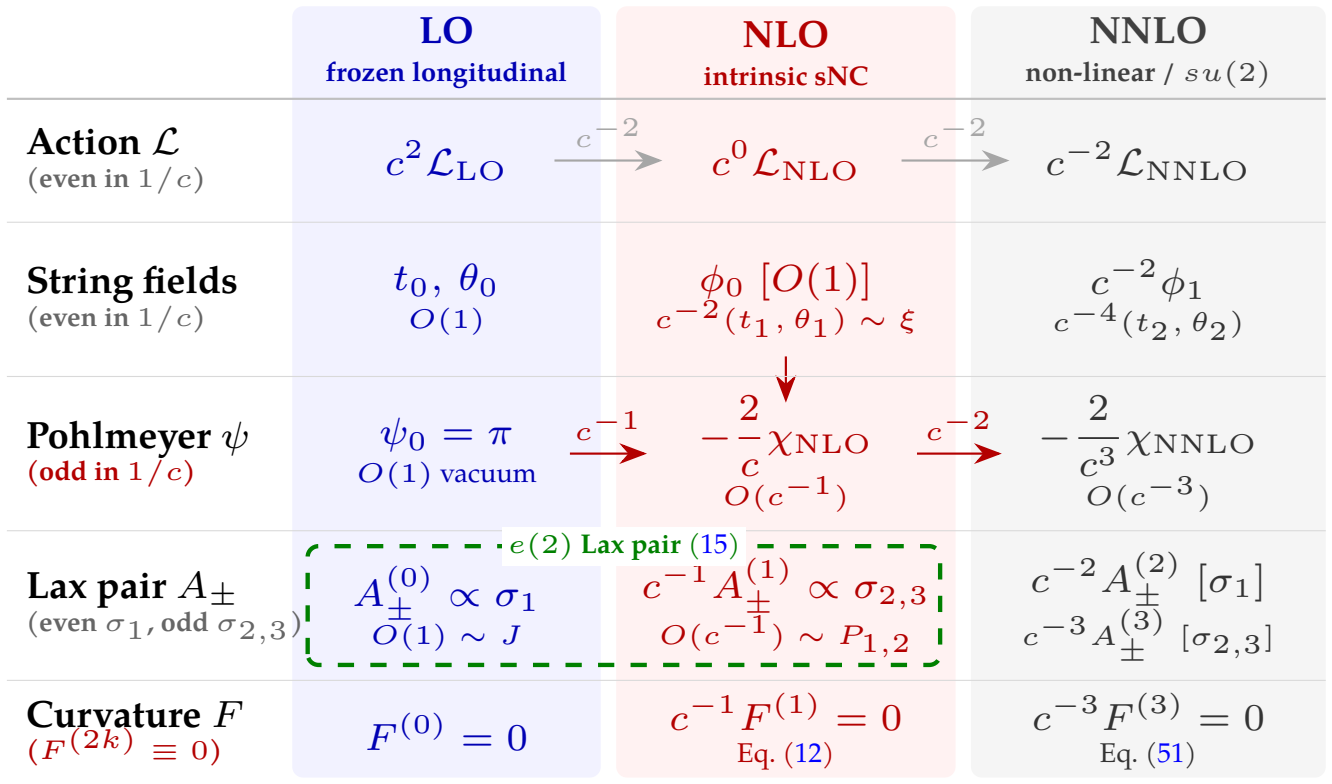

Note that $\chi$ is first order in $1/c$ within $\psi$, and at the same time NLO in the string expansion, since by \eqref{eq:chiexp} it is built from $\phi_0$. This first appears in $\mathcal L_{\rm NLO}$, and it obeys \eqref{eq:chiLO}, which is the NLO equation of motion, i.e. the NLO transverse equation \eqref{eq:NClin}. There is no contradiction: $\pi-\psi$ expands in odd powers of $1/c$, whereas the action runs in even powers. Similarly, $\chi_{\rm NNLO}$ enters $\psi$ at order $c^{-3}$ and is an NNLO object.\footnote{Only odd powers of $1/c$ appear in $\pi-\psi$. Since all fields and $\kappa_r$ are expanded in even powers of $1/c$, the normalised oriented area $\sin\psi=\sin\theta\,(\p_+\theta\,\p_-\phi-\p_-\theta\,\p_+\phi)/(c\,\p_+t\,\p_-t)$ is $1/c$ times an even series, while $\cos\psi$ is an even series. Since \eqref{eq:psidef} fixes $\psi$ only through $\cos\psi$, the sign of $\sin\psi$ is a convention, chosen here so that $\sin\psi$ equals the oriented area on the winding branch; on the momentum branch the same area equals $-\sin\psi$ with $\psi=2\chi_c/c$, cf.\ appendix~\ref{app:chirel}. Hence $\pi-\psi=\arcsin(\sin\psi)$ is odd. We have confirmed this on the exact rotating string.} The NR string sits close to the unstable vacuum because it is maximally stretched and almost at rest, and $\chi$ measures its small departure from rest. See figure~\ref{fig:orders} for a summary.

As an independent check, let us expand \eqref{eq:psidef} directly. We include the expansion of the normalisation $\p_+t\,\p_-t$ and identify $\xi_1=2(\theta_1'-\dot t_1)$, which follows from comparing the NLO Lagrangian with the intrinsic one (appendix~\ref{A3}), and which is how the subleading longitudinal fields reproduce the intrinsic multipliers. To order $1/c^2$, we find
\begin{equation}\label{eq:cospsi}
\begin{aligned}
1+\cos\psi&=\frac{\sin^2\vartheta\,(\dot\varphi^2-\varphi'^2)-\kappa\,\xi_1}{c^2\kappa^2}\\
&=\frac{2\sin^2\vartheta\,\dot\varphi^2}{c^2\kappa^2},
\end{aligned}
\end{equation}
where the second step uses the Virasoro solution $\xi_1=-\sin^2\vartheta(\dot\varphi^2+\varphi'^2)/\kappa$ of \eqref{eq:A:xisol}. Since $1+\cos\psi=2\sin^2\frac{\pi-\psi}{2}$, this reproduces \eqref{eq:chiexp}. The gradient contributions cancel between the transverse and multiplier terms, which is why $\chi$ does not see the twist. The same field descends from the relativistic normalised oriented area, as in appendix~\ref{app:chirel}.

Let us now take \eqref{eq:psiorders} only up to NLO. Substituting $\psi=\pi-2\chi/c$ into \eqref{eq:SG} gives an exact equation, before the full expansion is implemented:
\begin{equation}\label{eq:chiexact}
\begin{aligned}
\ddot\chi-\chi''&=\frac{c\,\kappa_r^2}{2}\sin\frac{2\chi}{c}\\
&=\kappa_r^2\Big(\chi-\frac{2\chi^3}{3c^2}\Big)+O(c^{-4}),
\end{aligned}
\end{equation}
which is the sine-Gordon equation about its maximum with coupling $2/c$. Only even powers of $1/c$ appear in \eqref{eq:chiexact}. At pure NLO it reduces to the intrinsic equation \eqref{eq:chiLO},
\begin{equation}
\ddot\chi_{\rm NLO}-\chi_{\rm NLO}''=\kappa^2\chi_{\rm NLO}.
\end{equation}
Including NNLO for $\chi$ and $\kappa_r=\kappa+h/c^2$, the order $c^{-2}$ of \eqref{eq:chiexact} gives the NNLO equation
\begin{equation}\label{eq:chiNNLOeq}
\ddot\chi_{\rm NNLO}-\chi_{\rm NNLO}''-\kappa^2\chi_{\rm NNLO}=2\kappa h\,\chi_{\rm NLO}-\frac{2\kappa^2}{3}\chi_{\rm NLO}^3 ,
\end{equation}
in which the NLO field acts as a source, through the frequency shift and the cubic coupling. The cubic term is an NNLO prediction, which we test on the rotating string ($\varphi=\omega\tau$), using $\kappa_r=\kappa+\omega^2/(4\kappa c^2)+O(c^{-4})$. The static form of \eqref{eq:chiexact} is solved by
\begin{equation}\label{eq:chinnlo}
\chi=\frac\omega\kappa\sin\vartheta-\frac{\omega^3}{12\kappa^3c^2}\sin^3\vartheta+O(c^{-4}).
\end{equation}
Its two terms are $\chi_{\rm NLO}$ and $c^{-2}\chi_{\rm NNLO}$, and the latter solves \eqref{eq:chiNNLOeq}, as verified. For this solution, the terms in $\sin\vartheta$ and in $\sin^3\vartheta$ determine the coefficient independently. The same expression follows from the expansion of the exact relativistic solution. This confirms the consistency of the expansion of the relativistic reduction at NNLO. It does not test the NNLO non-relativistic string theory directly, since we have not computed $\chi_{\rm NNLO}$ from the NNLO string fields.

\subsection{Winding sector: expansion of the Lax pair}\label{sec:laxexp}

Let us now expand the Lax pair \eqref{eq:Lax} and its curvature. Writing $\psi=\pi-2\chi/c$ and $\mu=\kappa_r$, both are series in $1/c$,
\begin{equation}\label{eq:Fexp}
\begin{gathered}
A_\pm=\sum_{n\ge0}c^{-n}A_\pm^{(n)},\qquad F=\sum_{n\ge0}c^{-n}F^{(n)},\\
F^{(n)}=\p_-A_+^{(n)}-\p_+A_-^{(n)}+\sum_{i+j=n}\big[A_+^{(i)},A_-^{(j)}\big].
\end{gathered}
\end{equation}
The commutator term couples different orders, and this is the origin of the mixing of orders in the reduced theory. Note also the parity of the series. Since $\psi-\pi$ is odd in $1/c$ expansion, the cosine in \eqref{eq:Lax} is even and the sine odd, so that
\begin{equation}
A_\pm^{(n)} \ \text{lies along}\ 
\begin{cases}
\sigma_1, & n \text{ even},\\[2pt]
\sigma_3\ (A_+) \ \text{or}\ \sigma_2\ (A_-), & n \text{ odd}.
\end{cases}
\end{equation}

 Moreover, the exact curvature $-\frac i2\big(\p_+\p_-\psi+\frac{\mu^2}{4}\sin\psi\big)\sigma_3$ is odd in $1/c$, because both $\p_+\p_-\psi$ and $\sin\psi$ are. Hence $F^{(2k)}=0$ identically, to all orders, and the reduced equations at successive orders come from $F^{(2k+1)}$. At the first two orders, with $\mu=\kappa$,
\begin{equation}\label{eq:Laxwind}
\begin{aligned}
A_+^{(0)}&=\frac{\lambda\kappa}{4}\sigma_1,& A_+^{(1)}&=i\,\p_+\chi\,\sigma_3,\\
A_-^{(0)}&=\frac{\kappa}{4\lambda}\sigma_1,& A_-^{(1)}&=-\frac{\kappa}{2\lambda}\chi\,\sigma_2 .
\end{aligned}
\end{equation}
Here $A^{(0)}_\pm$, which contains only the longitudinal frequency, is the LO Lax pair, and $A^{(1)}_\pm$, which contains the Pohlmeyer field and hence the NLO dynamics, is the NLO Lax pair. The LO pair is constant and abelian, so that $F^{(0)}=0$ identically, in agreement with the absence of LO dynamics. At the next order, \eqref{eq:Fexp} gives
\begin{equation}\label{eq:F1}
F^{(1)}=\p_-A_+^{(1)}-\p_+A_-^{(1)}+\big[A_+^{(0)},A_-^{(1)}\big]+\big[A_+^{(1)},A_-^{(0)}\big].
\end{equation}
The derivative terms give $i\,\p_+\p_-\chi\,\sigma_3+\frac{\kappa}{2\lambda}\p_+\chi\,\sigma_2$, while the two commutators give
\begin{equation}\label{eq:F1cross}
\big[A_+^{(0)},A_-^{(1)}\big]+\big[A_+^{(1)},A_-^{(0)}\big]=-\frac{i\kappa^2}{4}\chi\,\sigma_3-\frac{\kappa}{2\lambda}\p_+\chi\,\sigma_2 ,
\end{equation}
so that
\begin{equation}\label{eq:F1result}
F^{(1)}=i\Big(\p_+\p_-\chi-\frac{\kappa^2}{4}\chi\Big)\sigma_3 ,
\end{equation}
whose vanishing is \eqref{eq:chiLO}. The cross terms do two things: they cancel the $\sigma_2$ term, and they supply the tachyonic mass. Without them, flatness of the NLO pair alone would require both $\p_+\p_-\chi=0$ and $\p_+\chi=0$, which is neither the reduced equation nor consistent with it. Equivalently, \eqref{eq:F1} is the covariant flatness of the NLO pair with respect to the LO one,
\begin{equation}
D^{(0)}_-A_+^{(1)}-D^{(0)}_+A_-^{(1)}=0
\end{equation}
 with $D^{(0)}_\pm=\p_\pm-[A^{(0)}_\pm,\,\cdot\,]$, i.e.\ the linearisation of flatness about the flat LO connection.\footnote{Writing $A=A^{(0)}+\epsilon B$ with $\epsilon=1/c$, and using $F[A^{(0)}]=0$, the curvature is $F[A]=\epsilon\big(\p_-B_+-\p_+B_-+[A^{(0)}_+,B_-]+[B_+,A^{(0)}_-]\big)+O(\epsilon^2)$, whose bracket is $D^{(0)}_-B_+-D^{(0)}_+B_-$. The NLO pair is thus flat only together with its commutators with the LO pair, as for a gauge-field fluctuation about a fixed background.} Physically, the mass term of \eqref{eq:chiLO} is the coupling of the transverse NLO field to the longitudinal LO motion, encoded in $A^{(0)}$. This is why the Lax pair of the Pohlmeyer reduced theory, although its field $\chi$ is an NLO object, must contain both the LO and the NLO pairs.

The comparison with the intrinsic theory makes the same point from the other side. On the winding branch, the intrinsic Lagrangian \eqref{eq:NCL} is the NLO Lagrangian evaluated on the LO solution, with $\vartheta=\theta_0$ and the multipliers built from $t_1$ and $\theta_1$ (appendix~\ref{A3}). The intrinsic theory is therefore not the NLO theory alone, but the NLO theory on its frozen LO background, and its Lax pair must encode both. In \eqref{eq:LaxNC}, the LO data enter through the rotation $\p_\pm\vartheta\,J$, and the NLO data through the translations. It is thus the sum of the LO and NLO pairs, and not the NLO pair alone, that should be compared with the intrinsic pair. This is novel, compared to known constructions focusing only on field solutions at the action level.

This structure becomes transparent in contracted generators. Let us define the elements of $su(2)$
\begin{equation}\label{eq:contraction}
J=\frac i2\,\sigma_1,\qquad P_1=-\frac ic\,\sigma_2,\qquad P_2=\frac ic\,\sigma_3,
\end{equation}
which satisfy $[J,P_1]=P_2$ and $[J,P_2]=-P_1$ exactly, together with $[P_1,P_2]=4J/c^2$. These are the relations of $so(3)$ with the generators $P_{1,2}$ scaled by $1/c$, and they reduce to \eqref{eq:e2} as $c\to\infty$. The map \eqref{eq:contraction} is thus the In\"on\"u-Wigner contraction anticipated in section~\ref{sec:intrinsic}. Rewriting \eqref{eq:Laxwind} in terms of \eqref{eq:contraction} and renaming the spectral parameter as $\lambda\to i\lambda$, we find, term by term,
\begin{equation}\label{eq:LOlaxNLOlax}
\begin{aligned}
A_+^{(0)}&=\frac{\lambda\kappa}{2}J,& c^{-1}A_+^{(1)}&=\p_+\chi\,P_2,\\
A_-^{(0)}&=-\frac{\kappa}{2\lambda}J,& c^{-1}A_-^{(1)}&=-\frac{\kappa}{2\lambda}\chi\,P_1 .
\end{aligned}
\end{equation}
The explicit factor $1/c$ multiplying the NLO pair is absorbed into the generators $P_{1,2}$. Hence the LO pair supplies the rotational part $\p_\pm\vartheta\,J$ of \eqref{eq:LaxNC}, the longitudinal motion, and the NLO pair its translational part, the transverse dynamics. Their sum is precisely \eqref{eq:LaxNC}, as we have verified, so no gauge transformation is needed to relate the two, and the cross terms \eqref{eq:F1cross} become the commutators $[J,P_{1,2}]$ of \eqref{eq:e2}.

Put another way, on a sphere of radius $R=c$, as explained before, the longitudinal coordinate $\theta$ is an $O(1)$ angle along the meridian, whereas an $O(1)$ transverse displacement $y=\sin\theta\,\phi$ subtends a small angle $\sim y/c=O(1/c)$. In the relativistic $\mathfrak{su}(2)$ Pauli basis $\{\sigma_1,\sigma_2,\sigma_3\}$, every generator measures angles on $S^2$. Consequently, the meridian rotation along $\sigma_1$ is $O(1)$, while the transverse tilt along $\sigma_{2,3}$ is suppressed by $O(1/c)$. Writing \eqref{eq:Laxwind} with $\lambda\to i\lambda$, as in \eqref{eq:LOlaxNLOlax}, we find
\begin{equation}
\begin{aligned}
A_+ &= \underbrace{\frac{i\lambda\kappa}{4}\sigma_1}_{O(1)\text{ meridian angle}}
+ \underbrace{\frac{i}{c}\p_+\chi\,\sigma_3}_{O(1/c)\text{ transverse tilt}} + O(c^{-2}),\\[4pt]
A_- &= \underbrace{-\frac{i\kappa}{4\lambda}\sigma_1}_{O(1)\text{ meridian angle}}
+ \underbrace{\frac{i\kappa}{2c\lambda}\chi\,\sigma_2}_{O(1/c)\text{ transverse tilt}} + O(c^{-2}).
\end{aligned}
\end{equation}
In the contracted $\mathfrak{e}(2)$ basis \eqref{eq:contraction}, absorbing $1/c$ into $P_1=-\frac ic\sigma_2$ and $P_2=\frac ic\sigma_3$ converts these $O(1/c)$ angular tilts into $O(1)$ translations in the local tangent plane, so that both longitudinal and transverse terms enter the intrinsic Lax pair \eqref{eq:LaxNC} at $O(1)$.

For completeness, the content of the Lax pair becomes apparent at NNLO, where the reduced equation \eqref{eq:chiexact} is no longer linear. The details can be found in appendix~\ref{NNLOlax}, which also discusses how the $su(2)$ structure is restored order by order in $1/c^2$.

\subsection{Momentum sector}

The momentum sector, though NR in physics, has no intrinsic sNC description, since it lies on the degenerate branch of the LO Virasoro constraint.\footnote{\cite{HH1,HH2} assume that the pull-back of $\tau_{\mu\nu}$ to the worldsheet is Lorentzian, and their winding boundary conditions exclude this branch \cite{HH2}. Their analysis therefore does not settle whether it is a sector of the non-relativistic string or only a formal solution of the expanded equations. We treat it as the zero-winding sector of the expansion and use it mainly for comparison. Its status is an open question.} Its reduction follows from the expansion alone. Here $\dot\theta_0=\zeta$ and $\theta'=O(c^{-2})$, so the first relation in \eqref{eq:staticc} gives $\psi\to0$, and we define
\begin{equation}\label{eq:chicdef}
\chi_c\equiv\frac c2\,\psi=\frac{\sin\theta_0\,\phi_0'}{\zeta}+O(c^{-2}),
\end{equation}
the transverse gradient in units of $\zeta$.\footnote{This is the gradient invariant of section~\ref{phlf}, which does not obey a closed equation in the winding sector, but does on the momentum branch.} Here $\psi$ is expanded about the constant $0$, and $\chi_c$ is the NLO Pohlmeyer field of this sector. The string moves at nearly the speed of light and has small spatial extent, which is why it sits close to the stable vacuum. Substituting into \eqref{eq:SG}, with $\mu=\zeta_r$, gives
\begin{equation}\label{eq:chicLO}
\begin{gathered}
\ddot\chi_c-\chi_c''+\frac{c\,\zeta_r^2}{2}\sin\frac{2\chi_c}{c}=0,\\
\implies \ddot\chi_c-\chi_c''+\zeta^2\chi_c=0\quad\text{(NLO)}.
\end{gathered}
\end{equation}
This is a massive Klein-Gordon equation, the linearisation about the stable vacuum (appendix~\ref{app:vacua}). Its closed-string frequencies are $\omega_n=\sqrt{n^2+\zeta^2}$, which is the form of the transverse spectrum in a pp-wave \cite{BMN}, consistent with the light-cone structure of this sector (appendix~\ref{A3}). This structure is reminiscent of the NR strings obtained from near-BPS limits \cite{Harmark1}. The same equation follows from the NLO transverse equation $\p_\tau(\sin^2\theta_0\,\dot\phi_0)=\p_\sigma(\sin^2\theta_0\,\phi_0')$: with $y_c=\sin\theta_0\,\phi_0$, it becomes $\ddot y_c-y_c''+\zeta^2y_c=0$, and $\chi_c=y_c'/\zeta$.

The momentum sector also admits an $e(2)$ Lax pair of the same form as \eqref{eq:LaxNC},
\begin{equation}\label{eq:Laxmom}
A_+=\frac{\lambda\zeta}{2}J+\p_+\chi_c\,P_2,\qquad A_-=\frac{\zeta}{2\lambda}\big(J+\chi_c\,P_1\big),
\end{equation}
with curvature $(\p_+\p_-\chi_c+\frac{\zeta^2}{4}\chi_c)P_2$. Similarly, the LO and NLO Lax pairs obtained by expanding \eqref{eq:Lax} about $\psi=0$, rewritten with \eqref{eq:contraction} and $\lambda\to i\lambda$, coincide with \eqref{eq:Laxmom} up to the sign of $\chi_c$, which reflects $\chi_{\rm sNC}=-\chi_c$ on this sector.\footnote{Evaluating \eqref{eq:chicov} on this sector gives $-\chi_c$, which is a consistency check of the full setup.} The rotational part is again $\p_\pm\theta_0\,J$, now with $\p_+\theta_0=\p_-\theta_0$ in this branch. The two sectors therefore differ only in the longitudinal motion that enters the Lax pair, i.e.\ in the LO background with respect to which the NLO pair is covariantly flat.

The NLO energy is a light-cone Hamiltonian (see around \eqref{eq:A:lc}), which in terms of the Pohlmeyer field reads
\begin{equation}\label{eq:Emom}
E-P_\theta\Big|_{\rm NLO}=\frac{\pi T^2\zeta^2}{N_{\rm LO}}\oint\chi_c^2\,d\sigma ,
\end{equation}
with $N_{\rm LO}$ the LO oscillation number (appendix~\ref{app:puls}). A specific example to consider here is the pulsating string \cite{Minahan,Roy1} (figure~\ref{fig:strings}c), with ansatz $\phi_0=m\sigma$, having
\begin{equation}\label{pulsi}
\chi_c=\frac m\zeta\sin(\zeta\tau+\alpha_0),
\end{equation}
 and \eqref{eq:Emom} gives $2\pi^2T^2m^2\sin^2\theta_0/N_{\rm LO}$. Note that \eqref{eq:Emom} contains the instantaneous momentum $P_\theta$, whose NLO part oscillates. Its time average is $(\pi Tm)^2/N_{\rm LO}$, the constant energy correction of the pulsating string. As in the light-cone gauge, the NLO Virasoro constraints of this sector also impose a level-matching condition, $\oint\dot\phi_0\,\phi_0'\,d\sigma=0$, which restricts the allowed transverse data. The pulsating string satisfies it trivially, since $\dot\phi_0=0$, but a general superposition of modes does not, as we explain in appendix~\ref{app:levelmatch}.

\subsection{Winding-momentum duality}

The two sectors above are exchanged by the worldsheet exchange $\tau\leftrightarrow\sigma$, i.e.\ $\sigma^-\to-\sigma^-$ at fixed $\sigma^+$. Under it, the static relations \eqref{eq:staticc} exchange $\cos^2\frac\psi2$ and $\sin^2\frac\psi2$, so that
\begin{equation}\label{eq:duality}
\psi\;\to\;\pi-\psi,\qquad \chi=\frac{\sin\vartheta\,\dot\varphi}{\kappa}\;\leftrightarrow\;\chi_c=\frac{\sin\theta_0\,\phi_0'}{\zeta},
\end{equation}
which exchanges the unstable and stable vacua and the transverse velocity with the transverse gradient. The tachyonic equation \eqref{eq:chiLO} becomes the massive equation \eqref{eq:chicLO}, with $\kappa\to\zeta$. On the Lax pairs, $A_-\to-A_-$, which maps \eqref{eq:LaxNC} to \eqref{eq:Laxmom} with $\kappa\to\zeta$ and $\chi\to\chi_c$. At the level of the reduced theory, this is the worldsheet form of the exchange of winding and momentum, i.e.\ a kind of T-duality, consistent with the treatment of the zero-winding sector through a T-dual expansion in \cite{HH2}.

The two sectors are reminiscent of the two inequivalent Galilean limits of electrodynamics \cite{LeBellac:1973}, the electric and the magnetic limit, in which one of the two fields dominates and which are exchanged by duality. Here the limits are distinguished by which tangent of the string dominates: in the winding sector $|\p_\sigma\vec X|$ is maximal, the string is stretched and almost at rest, and the reduced field is a velocity. In the momentum sector $|\p_\tau\vec X|$ is maximal, the string moves at the speed of light, and the reduced field is a gradient. The duality \eqref{eq:duality} plays the role of the electric-magnetic exchange. We stress that this is an analogy at the level of the reduced theory: the assignment of the labels electric and magnetic is a convention, and no statement about a target-space electromagnetic duality is implied.

\subsection{Dictionary}

Let us now make the correspondence between NR strings and solutions of the Pohlmeyer reduced theory explicit, starting from the two simplest families, the rigidly rotating and the pulsating strings. Both have exact relativistic counterparts on the sphere of radius $c$. In static gauge, $t=\kappa_r\tau$, the rigidly rotating string is described by the ansatz $\phi_{\rm rel}=(\omega/c)\,\tau$ and $\theta=\theta(\sigma)$. The Virasoro constraints then reduce to $\theta'^2+(\omega/c)^2\sin^2\theta=\kappa_r^2$, which is the equation of a pendulum in $\sigma$, with solution
\begin{equation}\label{eq:rotexact}
\theta(\sigma)=\operatorname{am}\big(\kappa_r\sigma\,\big|\,k^2\big),\qquad k=\frac{\omega}{c\,\kappa_r},
\end{equation}
where $\operatorname{am}$ is the Jacobi amplitude. This is a closed string through both poles, rotating rigidly about the polar axis, whose profile along the meridian is slightly non-uniform. Closure on the meridian, $\theta(\sigma+2\pi)=\theta(\sigma)+2\pi\kappa$, fixes the frequency,
\begin{equation}\label{eq:rotclosure}
\kappa_r=\frac{2\kappa}{\pi}\,K(k^2)=\kappa+\frac{\omega^2}{4\kappa c^2}+O(c^{-4}),
\end{equation}
with $K$ the complete elliptic integral of the first kind. As $c\to\infty$, \eqref{eq:rotexact} reduces to $\theta=\kappa\sigma$, the NR rigid rotation, and \eqref{eq:rotclosure} reproduces independently the frequency shift obtained from the NLO Virasoro constraint in appendix~\ref{A3}.

Similarly, the pulsating string is described by $\phi_{\rm rel}=(m/c)\,\sigma$ and $\theta=\theta(\tau)$, with $t=\zeta_r\tau$. The Virasoro constraints now give a pendulum in $\tau$,
\begin{equation}\label{eq:pulsexact}
\theta(\tau)=\operatorname{am}\big(\zeta_r\tau\,\big|\,k_c^2\big),\qquad k_c=\frac{m}{c\,\zeta_r},
\end{equation}
which describes a circle of latitude, wound $m$ times in the azimuthal direction, that moves along the meridian and shrinks to a point each time it crosses a pole. This is the circular pulsating string of \cite{Minahan}, and as $c\to\infty$ it reduces to $\theta_0=\zeta\tau$, the NR pulsating string of the momentum sector. Note that $\phi_0=m\sigma$ closes for $m\in\mathbb Z$ in the squashed completion, whereas in the round one closure requires $m\in c\,\mathbb Z$. The local solution is the same in both.

The relativistic Pohlmeyer fields of these solutions follow from \eqref{eq:staticc}. For the rotating string $\dot\theta=0$, and the first relation gives $\cos\frac\psi2=k\sin\theta$. For the pulsating string $\theta'=0$, and the second relation gives $\sin\frac\psi2=k_c\sin\theta$. Hence
\begin{equation}\label{eq:exact}
\sin\frac{\pi-\psi}{2}=k\sin\theta,\qquad \sin\frac\psi2=k_c\sin\theta .
\end{equation}
Combined with \eqref{eq:rotexact} and \eqref{eq:pulsexact}, these give a static solution and a homogeneous solution of the sine-Gordon equation respectively, as we have verified. Let us locate them in the moduli space of elliptic solutions. The static equation $\psi''=\mu^2\sin\psi$ and the homogeneous equation $\ddot\psi=-\mu^2\sin\psi$ have the first integrals
\begin{equation}\label{eq:SGfirstint}
\frac12\psi'^2+\mu^2\cos\psi=E_{\rm SG},\qquad \frac12\dot\psi^2-\mu^2\cos\psi=E_{\rm SG},
\end{equation}
and in both cases the vacuum about which we expand has $E_{\rm SG}=-\mu^2$. Evaluating \eqref{eq:SGfirstint} at a turning point of $\psi$, where $\sin\theta=1$, and using \eqref{eq:exact} with $\mu=\kappa_r$ and $\mu=\zeta_r$, we find
\begin{equation}\label{eq:Eshift}
E_{\rm SG}+\mu^2=\frac{2\omega^2}{c^2}~\text{(rotating)},~ E_{\rm SG}+\mu^2=\frac{2m^2}{c^2}~\text{(pulsating)},
\end{equation}
exactly. The NR strings are therefore elliptic solutions whose first integral approaches the double root $E_{\rm SG}=-\mu^2$ as $c^{-2}$. Note that the elliptic solutions of the sine-Gordon equation form a family labelled by the first integral $E_{\rm SG}$, which is the moduli space studied in \cite{KMP1}. The value $E_{\rm SG}=-\mu^2$, at which two roots of the underlying elliptic polynomial coincide, corresponds to the vacuum itself, and solutions with $E_{\rm SG}$ close to it are small oscillations about the vacuum. The NR strings thus occupy this vacuum corner of the moduli space, and their sine-Gordon counterparts are small oscillations about the two vacua.

The remaining entries of the dictionary follow from the analysis of the previous sections. The string at rest along a meridian is the vacuum $\psi=\pi$ itself, and its fluctuations are the segment oscillators \eqref{eq:chimodes}, with frequencies $\omega_n=\kappa\sqrt{n^2-1}$, of which the rigid rotation is the static mode $n=1$. The point-like string moving at the speed of light is the vacuum $\psi=0$, the NR analogue of the BMN particle, and its fluctuations on the closed string have the pp-wave frequencies $\omega_n=\sqrt{n^2+\zeta^2}$, with the pulsating string as the homogeneous mode. Twisted rotating strings have the same Pohlmeyer field as the untwisted one, as shown in section~\ref{sec:intrinsic}, and they are singular at the poles. Kinks and giant magnons have no counterpart, since they reach the other vacuum (see section~\ref{sec:kinks}). Table~\ref{tab:dict} collects the dictionary, together with the embeddings in the ansatz commonly used for classical strings, so that each entry can be recognised directly.

\begin{table*}[t]
\centering\footnotesize
\renewcommand{\arraystretch}{1.25}
\begin{tabular}{p{0.17\textwidth}|p{0.36\textwidth}|p{0.36\textwidth}}
NR string & Embedding (static gauge) & Reduced counterpart\\\hline
At rest on a meridian & $t=\kappa\tau$, $\vartheta=\kappa\sigma+\sigma_0$, $\varphi={\rm const}$ & $\chi=0$, i.e.\ $\psi=\pi$; static great circle\\
Rigid rotation & as above, with $\varphi=\omega\tau$ & mode $n=1$ of \eqref{eq:chimodes}, $\chi=\frac\omega\kappa\sin\vartheta$; exact solution \eqref{eq:rotexact}\\
Segment oscillators & as above, with $\sin\vartheta\,\varphi\propto\cos(\omega_n\tau+\beta_n)\sin n(\vartheta-j\pi)$ & Dirichlet modes \eqref{eq:chimodes}\\
Twisted rotation & as above, with $\varphi=\omega\tau+f(\sigma)$, $\sin^2\vartheta\,f'=v\neq0$ & same $\chi$ as $v=0$; singular at the poles\\\hline
Point-like & $t_0=\theta_0=\zeta\tau+\alpha_0$, $\phi_0={\rm const}$ & $\chi_c=0$, i.e.\ $\psi=0$; BMN particle\\
Pulsating string & $t_0=\theta_0=\zeta\tau+\alpha_0$, $\phi_0=m\sigma$ & homogeneous $\chi_c=\frac m\zeta\sin(\zeta\tau+\alpha_0)$; exact solution \eqref{eq:pulsexact}\\
Small breathers (on the line) & momentum sector, $\epsilon=\alpha/c$ in \eqref{eq:breather} & pulsating string, $m=2\alpha\zeta$ (section~\ref{sec:kinks})\\\hline
Kinks, giant magnons & none & reach the other vacuum (section~\ref{sec:kinks})
\end{tabular}
\caption{Dictionary between NR strings, their embeddings, and their Pohlmeyer counterparts. The upper block is the winding sector, described intrinsically in terms of $(t,\vartheta,\varphi)$, and the middle block is the momentum sector, described in terms of the LO and NLO fields of the expansion.}
\label{tab:dict}
\end{table*}
\section{Kinks and breathers}\label{sec:kinks}

The sine-Gordon equation has soliton solutions, and it is natural to ask which of them survive the NR limit. Let us recall that the relativistic theory has two kinds of solitons \cite{KMP1}: kinks, which interpolate between neighbouring stable vacua and are topological, and breathers, which are bound states of a kink and an antikink. We shall see that kinks decouple from the NR expansion, while small breathers survive and reduce to the pulsating string. Spikes, which are features of the string embedding rather than solitons of the reduced field, are discussed separately. Table~\ref{tab:solitons} summarises the $1/c$ scaling, sector assignment, and non-relativistic fate of each class of solutions.

\subsection{Why kinks disappear}

\begin{figure*}[t]
\centering
\includegraphics[width=\textwidth]{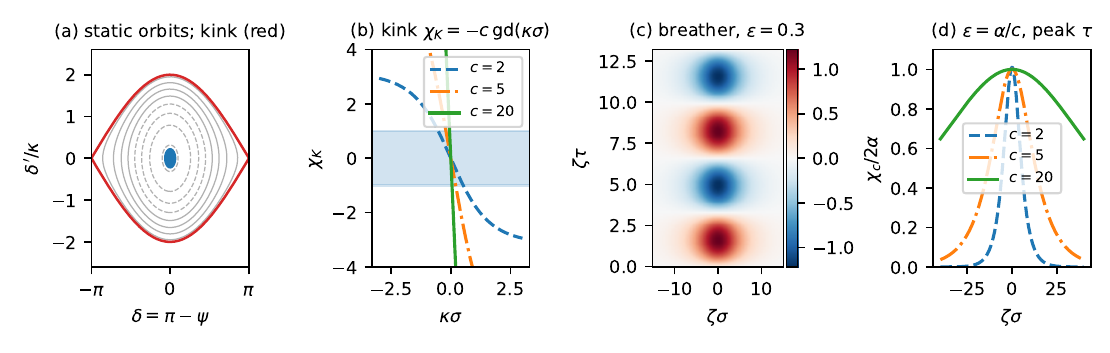}
\caption{Kinks and breathers. (a) Static orbits of the reduced equation about $\psi=\pi$; the NR solutions lie in the small shaded region, and the kink is the separatrix (red). (b) The kink \eqref{eq:kink} for increasing $c$, which leaves the band $|\chi|\le1$ within a core of width of order $1/(c\kappa)$. (c) The breather \eqref{eq:breather} for $\epsilon=0.3$. (d) The rescaled breather at its maximum, for $\epsilon=\alpha/c$ with $\alpha=1/2$, broadening into the homogeneous pulsating string as $c$ grows.}
\label{fig:kinks}
\end{figure*}

In static gauge, with $\mu=\kappa_r$, the single-kink solution of \eqref{eq:SG} is
\begin{equation}\label{eq:kinkpsi}
\psi_K(\sigma)=4\arctan e^{\kappa_r\sigma},
\end{equation}
which interpolates between the stable vacua $\psi=0$ ($\sigma\to-\infty$) and $\psi=2\pi$ ($\sigma\to+\infty$), passing through the unstable vacuum $\psi=\pi$ at $\sigma=0$ (appendix~\ref{app:vacua}). In terms of the winding-sector parent field $\chi=\frac c2(\pi-\psi)$, \eqref{eq:kinkpsi} becomes
\begin{equation}\label{eq:kink}
\chi_K(\sigma)=-c\,\operatorname{gd}(\kappa_r\sigma),\qquad \operatorname{gd}(x)=2\arctan e^{x}-\frac\pi2 ,
\end{equation}
where $\operatorname{gd}$ is the Gudermannian function.\footnote{The Gudermannian function, $\operatorname{gd}(x)=2\arctan e^x-\frac\pi2=\arctan(\sinh x)$, is odd and monotonic, behaves as $\operatorname{gd}(x)\simeq x$ for $|x|\ll1$, and tends to $\pm\frac\pi2$ as $x\to\pm\infty$. It satisfies $\sin\operatorname{gd}(x)=\tanh x$ and $\cos\operatorname{gd}(x)=\operatorname{sech}x$.} Equation \eqref{eq:kink} solves the static form of \eqref{eq:chiexact} exactly and forms the separatrix of the phase portrait in figure~\ref{fig:kinks}a. At fixed $c$, its asymptotic and central behaviours are
\begin{equation}\label{eq:kinklimits}
\begin{gathered}
\chi_K\to\mp\frac{\pi c}{2}\quad(\sigma\to\pm\infty),\\
\chi_K\simeq-c\,\kappa_r\sigma\quad(|\sigma|\ll\kappa_r^{-1}),
\end{gathered}
\end{equation}
so that $\chi_K$ interpolates between the stable vacua $\chi=\mp\frac{\pi c}{2}$ with central slope $-c\,\kappa_r$.

When $c\to\infty$ at fixed $\sigma$, the kink width $\sim\kappa_r^{-1}$ remains fixed while its step height $\pi c$ diverges, so $\chi_K(\sigma)$ has no finite pointwise limit. The NR regime $|\chi|\lesssim 1$ holds only inside the narrow core
\begin{equation}
|\sigma|\lesssim\frac{1}{c\,\kappa_r}\;\longrightarrow\;0,
\end{equation}
which shrinks to the origin as $c\to\infty$ (figure~\ref{fig:kinks}b). Equivalently, in the phase portrait of figure~\ref{fig:kinks}a, NR solutions are small orbits of radius $|\pi-\psi|\sim O(c^{-1})$ about the unstable fixed point $\psi=\pi$, whereas the kink separatrix lies at a distance of $O(1)$.

The worldsheet energetics of the reduced theory shows the same decoupling. Substituting $\psi=\pi-2\chi/c$ into the sine-Gordon Lagrangian \eqref{eq:Vsg} with $\mu=\kappa_r$, and multiplying by $c^2/4$ so that $\chi$ has a canonically normalised kinetic term, gives $\mathcal L_\chi=\frac{c^2}{4}\mathcal L_{\rm SG}=\frac12(\dot\chi^2-\chi'^2)-U(\chi)$. Its static worldsheet Hamiltonian density (as distinct from the spacetime string energy \eqref{eq:Ewind}) is
\begin{equation}\label{eq:HSG}
\mathcal H=\frac12\chi'^2+U(\chi),\qquad U(\chi)=\frac{c^2\kappa_r^2}{4}\Big(1+\cos\frac{2\chi}{c}\Big).
\end{equation}
The potential $U(\chi)$ vanishes at the stable vacua $\chi=\pm\frac{\pi c}{2}$ ($\psi=0,2\pi$) and has a maximum $U(0)=\frac12 c^2\kappa_r^2$ at the unstable vacuum $\chi=0$ ($\psi=\pi$). For NR configurations, where $\chi\sim O(1)$ stays near the maximum, $U(\chi)$ can be expanded in powers of $1/c^2$:
\begin{equation}\label{eq:Uexp}
U(\chi)=\frac{c^2\kappa_r^2}{2}-\frac{\kappa_r^2}{2}\chi^2+\frac{\kappa_r^2}{6c^2}\chi^4+O(c^{-4}),
\end{equation}
exhibiting the constant vacuum offset $U(0)$, the tachyonic mass term of \eqref{eq:chiLO}, and the quartic vertex behind the cubic term of \eqref{eq:chiexact}. Measured relative to $U(0)$, an NR excitation therefore carries reduced worldsheet energy density
\begin{equation}
\mathcal H-U(0)=\frac12\chi'^2-\frac{\kappa_r^2}{2}\chi^2+O(c^{-2})\sim O(1).
\end{equation}
For the kink \eqref{eq:kink}, by contrast, $\chi_K\sim O(c)$ reaches the stable vacua $\chi=\mp\frac{\pi c}{2}$ where $U=0$, so the cosine in \eqref{eq:HSG} cannot be expanded. Using $\chi_K'=-c\kappa_r\operatorname{sech}(\kappa_r\sigma)$ and $1+\cos(2\chi_K/c)=2\operatorname{sech}^2(\kappa_r\sigma)$, the reduced kink energy is
\begin{align}
E_{\chi,\text{kink}}&=\int_{-\infty}^{\infty}\mathcal H(\chi_K)\,d\sigma\nonumber \\
&=c^2\kappa_r^2\int_{-\infty}^{\infty}\operatorname{sech}^2(\kappa_r\sigma)\,d\sigma=2\kappa_r c^2,
\end{align}
which is the standard soliton mass $8\mu/\beta^2$ at coupling $\beta=2/c$ (appendix~\ref{app:vacua}). The ratio of the kink energy to that of an NR excitation of $\chi$ is therefore of order $c^2$, independent of whether one normalises the reduced theory in terms of $\chi$ or $\psi$.

The same $O(c^2)$ separation appears in the spacetime string energy $E$. On a sphere of radius $R=c$, the energy of a winding string expands as
\begin{equation}
E=c^2E_{\rm LO}+E_{\rm NLO}+O(c^{-2}),\qquad c^2E_{\rm LO}=2\pi T\kappa\,c^2,
\end{equation}
where $c^2E_{\rm LO}$ is the rest energy of the stretched meridian string and $E_{\rm NLO}\sim O(1)$ in \eqref{eq:Eintr} is the energy of the NR excitations. Relativistic giant magnons on $\mathbb R\times S^2$, which are the string counterparts of sine-Gordon kinks, obey the dispersion relation \cite{HM}
\begin{equation}
E-J=2Tc^2\Big|\sin\frac p2\Big|\sim O(c^2).
\end{equation}
Note here that the 't Hooft coupling $\sqrt{\lambda} = \frac{R^2}{\alpha'} =2\pi  T R^2$.
Their excitation energy (with momenta $O(1)$) is of the same order $O(c^2)$ as the LO rest energy, and parametrically larger than the NLO energies of the NR theory.

Kinks and giant magnons are thus heavy, non-perturbative states that decouple as $c\to\infty$. At the level of the string embedding, a kink is centred on $\psi=\pi$ at $\sigma=0$ but approaches $\psi=0,2\pi$ as $\sigma\to\mp\infty$, where \eqref{eq:staticc} gives
\begin{equation}
c^2\dot\theta^2+\sin^2\theta\,\dot\phi^2\to c^2\kappa_r^2,\qquad c^2\theta'^2+\sin^2\theta\,\phi'^2\to 0.
\end{equation}
In its tails the string therefore contracts and moves at the speed of light, which is incompatible with the winding-sector expansion $\theta'=\kappa+O(c^{-2})$ and $\dot\theta=O(c^{-2})$. In the intrinsic sNC theory this absence is immediate: \eqref{eq:chiLO} is linear and confined near $\psi=\pi$, so there are no other vacua and no topological sectors. Note that kink-like profiles of the longitudinal coordinates were found in non-relativistic $AdS_5\times S^5$ in a different residual gauge, in which the multipliers are constant \cite{FN2}. These require a compact time coordinate, and they are not kinks of the reduced field. Our statement concerns the latter, in static gauge.

\subsection{Spikes}

A kink is a soliton of the reduced field: it interpolates between distinct
stable vacua of $\psi$. A spike is instead a local feature of the string
embedding. In static gauge, define the squared length of the spatial string
tangent by
\begin{equation}
\ell_\sigma^2\equiv
c^2\theta'^2+\sin^2\theta\,\phi'^2 .
\end{equation}
The second Virasoro relation in \eqref{eq:staticc} gives
\begin{equation}\label{eq:spikecondition}
\ell_\sigma^2=c^2\mu^2\sin^2\frac{\psi}{2},
\qquad
\ell_\sigma=0\ \Longleftrightarrow\ \psi\in2\pi\mathbb Z ,
\end{equation}
where $\mu=\kappa_r$ in the winding sector and $\mu=\zeta_r$ in the
momentum sector. An isolated zero of $\ell_\sigma$ is the local condition
for a spike; whether the image has a cusp also depends on its local shape.
The kink \eqref{eq:kinkpsi} does not have such a zero at finite $\sigma$:
it passes through $\psi=\pi$ at its centre and reaches $\psi=0,2\pi$ only
asymptotically.

In the sNC winding sector, again using $\psi=\pi-\frac{2\chi}{c}+O(c^{-3}),~ \chi=O(1)$,
\begin{equation}
\ell_\sigma^2
=c^2\kappa_r^2\cos^2\!\left(\frac{\chi}{c}+O(c^{-3})\right)
=c^2\kappa^2+O(1).
\end{equation}
Thus the spatial tangent remains nonzero in the controlled winding
expansion. Reaching $\ell_\sigma=0$ would require $\chi=O(c)$, outside
the regime in which the expansion about $\psi=\pi$ is valid.

In the momentum sector, by contrast,
\begin{equation}
\psi=\frac{2\chi_c}{c}+O(c^{-3}),\qquad \chi_c=O(1),
\end{equation}
and hence
\begin{equation}
\ell_\sigma^2
=c^2\zeta_r^2\sin^2\!\left(\frac{\chi_c}{c}+O(c^{-3})\right)
=\zeta^2\chi_c^2+O(c^{-2}).
\end{equation}
For the pointlike BMN solution the tangent vanishes everywhere, not at an
isolated point. For the homogeneous pulsating solution,
$\phi_0=m\sigma$ and
$\chi_c=(m/\zeta)\sin(\zeta\tau+\alpha_0)$. Thus $\ell_\sigma=0$ when the hoop reaches a pole,
$\sin\theta=0$; at that time the entire hoop collapses to a point. This is
not an isolated spike. For general momentum-sector fluctuations, however,
zeros of the leading field $\chi_c$ do not by themselves establish an exact
zero of $\ell_\sigma$: subleading terms can contribute. We therefore make
no general claim about isolated spikes in arbitrary momentum-sector
solutions.

Outside static gauge, the longitudinal chirality conditions give
$X^+=F(\sigma^+)$ and $X^-=G(\sigma^-)$. With
$t=(X^++X^-)/2$ and $\theta=(X^+-X^-)/2$, the pulled-back longitudinal
metric is
\begin{equation}
\tau(\p_+X,\p_-X)=-\frac12 F'G'.
\end{equation}
A zero of $F'$ or $G'$ makes this longitudinal pullback degenerate, but
does not by itself make the full spatial tangent vanish. To see the additional condition, consider a regular slice of constant target time, $dt = t_\sigma d\sigma + t_\tau d\tau = 0$ with $t_\tau\ne0$, along which the spatial derivative is
\begin{equation}
\mathcal D_\sigma\equiv\left(\frac{\p}{\p\sigma}\right)_t =
\p_\sigma-\frac{t_\sigma}{t_\tau}\p_\tau .
\end{equation}Then
\begin{equation}
\mathcal D_\sigma\theta=\frac{F'G'}{t_\tau},\qquad
\left|\mathcal D_\sigma\vec X\right|^2
=c^2\big(\mathcal D_\sigma\theta\big)^2
+\sin^2\theta\big(\mathcal D_\sigma\phi\big)^2 .
\end{equation}
A zero of $F'G'$ removes the longitudinal contribution, but a spike also
requires the transverse contribution to vanish at the same isolated
point. The longitudinal profile alone can fold back on itself i.e. it does
not determine whether the full string has a cusp. Away from static gauge,
that question depends on the transverse dynamics as well.

\subsection{Breathers and the pulsating string}

Whereas kinks interpolate between distinct vacua and have a fixed topological step $\Delta\psi=2\pi$, the stable vacuum $\psi=0$ also supports breathers: oscillatory kink-antikink bound states that return to $\psi=0$ as $\sigma\to\pm\infty$ and depend on a continuous amplitude parameter.

In static gauge with $\mu=\zeta_r$, the sine-Gordon equation \eqref{eq:SG} reads
\begin{equation}
\ddot\psi-\psi''+\zeta_r^2\sin\psi=0.
\end{equation}
Substituting the separable ansatz $\psi(\sigma,\tau)=4\arctan[F(\tau)/G(\sigma)]$ with $F(\tau)=A\sin(\varpi\tau)$ and $G(\sigma)=\cosh(q\sigma)$ yields a solution on the line $\sigma\in\mathbb R$ provided
\begin{equation}\label{eq:lamb}
\varpi^2+q^2=\zeta_r^2,\qquad A=\frac q\varpi .
\end{equation}
Parametrising $q=\epsilon\zeta_r$ with $0<\epsilon<1$ fixes $\varpi=\sqrt{1-\epsilon^2}\,\zeta_r$ and $A=\epsilon/\sqrt{1-\epsilon^2}$, giving the breather about $\psi=0$,
\begin{equation}\label{eq:breather}
\psi_B(\sigma,\tau)=4\arctan\left[\frac{\epsilon\,\sin\big(\sqrt{1-\epsilon^2}\,\zeta_r\tau\big)}{\sqrt{1-\epsilon^2}\,\cosh(\epsilon\,\zeta_r\sigma)}\right].
\end{equation}
For any $0<\epsilon<1$, \eqref{eq:breather} is an exact relativistic solution oscillating at frequency $\sqrt{1-\epsilon^2}\,\zeta_r$ inside a spatial envelope of width $(\epsilon\zeta_r)^{-1}$ and peak amplitude $4\arctan(\epsilon/\sqrt{1-\epsilon^2})$ (figure~\ref{fig:kinks}c). When $\epsilon=O(1)$, $\psi_B$ reaches values of $O(1)$ and lies outside the NR regime.

Because $\epsilon$ is continuous, however, the amplitude can be scaled with $1/c$. Setting $\epsilon=\alpha/c$ with $\alpha$ fixed keeps $\psi_B=O(c^{-1})$ near the stable vacuum $\psi=0$, placing the breather in the momentum sector. Expanding the rescaled Pohlmeyer field $\chi_c=\frac c2\psi_B$ via $\arctan x=x+O(x^3)$ gives
\begin{equation}\label{eq:breatherexp}
\chi_c(\sigma,\tau)=\frac{2\alpha}{\sqrt{1-\alpha^2/c^2}}\,\frac{\sin\big(\sqrt{1-\alpha^2/c^2}\,\zeta_r\tau\big)}{\cosh(\alpha\zeta_r\sigma/c)}+O(c^{-2}).
\end{equation}
Taking $c\to\infty$ at fixed $\sigma$ and $\tau$, with $\zeta_r\to\zeta$, sends the frequency to $\zeta$ and $\cosh(\alpha\zeta_r\sigma/c)\to 1$, so that
\begin{equation}\label{eq:breatherlim}
\chi_c(\sigma,\tau)\;\longrightarrow\;2\alpha\sin(\zeta\tau).
\end{equation}
Comparing with \eqref{pulsi} shows that \eqref{eq:breatherlim} is the homogeneous pulsating string with $\alpha_0=0$ and transverse winding $m=2\alpha\zeta$. At finite $c$, \eqref{eq:breather} is a localised profile on $\sigma\in\mathbb R$ rather than a $2\pi$-periodic closed-string solution. As $c\to\infty$, its envelope width $c/(\alpha\zeta)$ diverges compared with the string length $2\pi$ (figure~\ref{fig:kinks}d); on the closed string only the flat central core $|\sigma|\ll c/(\alpha\zeta)$ remains visible, producing the spatially homogeneous pulsating hoop \eqref{eq:breatherlim}.

\subsection{Excluded modes}

Finally, the relativistic linearisation about $\psi=\pi$ has unstable periodic modes with $p<\kappa$, of frequency squared $p^2-\kappa^2$, which the Dirichlet condition excludes. By \eqref{eq:staticc}, a mode with $\chi\neq0$ at a pole requires a longitudinal velocity of order $1/c$, which is non-analytic in $1/c^2$ and outside the expansion. We expect these modes to describe the unwinding of the string, which is not topologically protected on the sphere, but we have not verified this. 

\begin{table}[t]
\centering
\caption{Scaling and $c\to\infty$ fate of configurations in the winding (W, $\psi_0=\pi$) and momentum (M, $\psi_0=0$) sectors.}
\label{tab:solitons}
\renewcommand{\arraystretch}{1.2}
\setlength{\tabcolsep}{3pt}
\resizebox{\columnwidth}{!}{%
\begin{tabular}{lllll}
\hline\hline
Configuration & Sec. & $\chi,\chi_c$ & $\Delta E$ & Fate as $c\to\infty$ \\
\hline
Kink / magnon & $0\to 2\pi$ & $O(c)$ & $O(c^2)$ & Decouples (core $\sim c^{-1}$) \\
Spike ($\ell_\sigma=0$) & W & $O(c)$ & --- & Excluded ($\ell_\sigma\sim c\kappa\ne 0$) \\
Spike ($\ell_\sigma=0$) & M & $0$ & $O(1)$ & Global in BMN / hoop \\
Breather ($\epsilon\sim 1$) & M & $O(c)$ & $O(c^2)$ & Decouples \\
Breather ($\epsilon=\alpha/c$) & M & $O(1)$ & $O(1)$ & Hoop (NLO) \\
Rotating string & W & $O(1)$ & $O(1)$ & Survives at NLO \\
Pulsating hoop & M & $O(1)$ & $O(1)$ & Survives at NLO \\
$n=0$ tachyon & W & $O(1)$ & --- & Excluded by pole BCs \\
\hline\hline
\end{tabular}%
}
\end{table}

\section{Conclusions}

In this work, inspired by the powerful techniques of integrability in relativistic strings in $AdS\times S$, we have formulated a Pohlmeyer-type reduction for classical non-relativistic strings on the string Newton-Cartan background $\mathbb R\times S^2$, both intrinsically and through the $1/c^2$ expansion of the relativistic string. The reduced field is the normalised oriented area \eqref{eq:chicov}, which obeys a tachyonic Klein-Gordon equation with Dirichlet walls at the poles, and the two descriptions agree on the winding sector.

Our results can be summarised as follows. First, the Pohlmeyer field of an sNC string is defined intrinsically, from the longitudinal and transverse projections of the tangents, without reference to a relativistic parent. Second, the $e(2)$ Lax pair \eqref{eq:LaxNC} is the Maurer-Cartan form of an intrinsic moving frame, and its spectral parameter is the rapidity of a worldsheet boost. Its monodromy yields conserved charges only at resonant values of the spectral parameter, where they reproduce the oscillator invariants of the reduced theory. Third, non-relativistic strings are identified as expansions about the two vacua of the sine-Gordon equation. The winding sector sits at the unstable vacuum and the momentum sector at the stable one, and the two are exchanged by the worldsheet exchange $\tau\leftrightarrow\sigma$, in a way reminiscent of the electric and magnetic Galilean limits. Fourth, the intrinsic Lax pair is the sum of the LO and NLO sine-Gordon pairs under an In\"on\"u-Wigner contraction. The mixing of orders follows from covariant flatness: the mass term of the reduced equation is the coupling of the transverse NLO field to the frozen LO motion. Finally, the reduction gives a sharp criterion for which relativistic configurations survive the limit, table~\ref{tab:dict}. Kinks, giant magnons and spikes in static gauge are heavy and decouple, while small breathers on the line reduce to the pulsating string.

These results open several directions of investigation. The most immediate is to relate the NNLO Pohlmeyer field to the NNLO string fields, which would turn the consistency check of section~\ref{sec:expansion} into a test of the NNLO non-relativistic string theory, and to compute the monodromy of the NNLO Lax pair, whose charges are sensitive to the non-linear structure. A second direction is the momentum sector, which lies on the branch excluded by a Lorentzian pull-back. Whether it describes genuine non-relativistic strings, and whether the winding-momentum exchange extends beyond the reduced theory to a duality of the full expansion, deserves a systematic study. A third concerns configurations outside the static gauge, where degenerate longitudinal maps combined with stationary transverse excitations may produce spikes, and where the transverse dynamics is no longer free. The singular Sturm-Liouville problem behind the Dirichlet walls, and its consequences for the spectrum, will be discussed elsewhere \cite{Banerjee:Upcoming}. It would also be valuable to relate \eqref{eq:LaxNC} to the sigma-model Lax pairs of \cite{Kluson,RoyLax,Fontanella:2020eje,Fontanella:2022fjd}, and its monodromy to the non-relativistic string monodromies of \cite{Fontanella:2022wfj}. Beyond the sphere, backgrounds with a curved or torsional longitudinal geometry, such as the $AdS_2$ longitudinal space of non-relativistic $AdS_5\times S^5$ \cite{FN2}, require a modified normalisation of the Pohlmeyer field. Extending the reduction to these cases, and to other symmetric spaces, would test how much of the present structure, in particular the contraction of the reduced Lax algebra and the role of the two vacua, is generic to non-relativistic strings. We hope to come back to some of these problems in the near future. 

\section*{Acknowledgements}
ABan was supported in part by an OPERA grant and a seed grant NFSG/PIL/2023/P3816 from BITS-Pilani, and further an early career research grant ANRF/ECRG/2024/002604/PMS from ANRF India. He also acknowledges financial support from the Asia Pacific Center for Theoretical Physics (APCTP) via an Associate Fellowship.

\subsection*{Data/Code Availability}
No data or proprietary code was used in this work.

\subsection*{AI usage}
Free public LLMs (ChatGPT 6 and Claude Sonnet 5.5) were used to draw the figures and to augment the numerical plots. They were also used as adversarial referees to help clarify certain parts of the manuscript. 

\appendix
\section{Vacua of the sine-Gordon equation}\label{app:vacua}

In static gauge, the sine-Gordon equation $\ddot\psi-\psi''+\mu^2\sin\psi=0$ follows from the Lagrangian
\begin{equation}\label{eq:Vsg}
\mathcal L_{\rm SG}=\frac12\big(\dot\psi^2-\psi'^2\big)-\mu^2\big(1-\cos\psi\big).
\end{equation}
Its constant solutions are the extrema of the potential $V=\mu^2(1-\cos\psi)$: the minima $\psi=2\pi n$, with $V=0$, and the maxima $\psi=\pi+2\pi n$, with $V=2\mu^2$. Linearising about them, $\psi=\psi_*+\delta$, gives
\begin{equation}\label{eq:vaclin}
\begin{aligned}
\psi_*=2\pi n:&\quad \ddot\delta-\delta''+\mu^2\delta=0,\quad \omega^2=p^2+\mu^2,\\
\psi_*=\pi+2\pi n:&\quad \ddot\delta-\delta''-\mu^2\delta=0,\quad \omega^2=p^2-\mu^2 .
\end{aligned}
\end{equation}
The minima are therefore stable vacua with massive fluctuations, while the maxima are unstable, with tachyonic fluctuations that grow for $p^2<\mu^2$. By \eqref{eq:static}, the stable vacuum $\psi=0$ has $|\p_\tau\vec X|=R\mu$ and $|\p_\sigma\vec X|=0$: the string is point-like and moves at the speed of light, which is the BMN particle. The unstable vacuum $\psi=\pi$ has $|\p_\tau\vec X|=0$ and $|\p_\sigma\vec X|=R\mu$: the string is at rest and maximally stretched along a great circle. A kink, $\psi_K=4\arctan e^{\mu\sigma}$, interpolates between neighbouring stable vacua, $\psi=0$ and $\psi=2\pi$, passing through the unstable one at its centre, with energy $8\mu$ in the normalisation of \eqref{eq:Vsg}. In a theory with coupling $\beta$, obtained by rescaling the field as $\psi=\beta\phi$, the Lagrangian becomes
\begin{equation}\label{eq:SGbeta}
\mathcal L=\frac12\big(\dot\phi^2-\phi'^2\big)-\frac{\mu^2}{\beta^2}\big(1-\cos\beta\phi\big),
\end{equation}
and the kink mass becomes $8\mu/\beta^2$. At weak coupling the kink is therefore heavy compared with the elementary excitations, whose mass is $\mu$, and it is a non-perturbative state.

In the non-relativistic expansion, the winding sector is the expansion about the unstable vacuum, and its reduced equation \eqref{eq:chiLO} is the tachyonic linearisation \eqref{eq:vaclin} with $\mu=\kappa$, stabilised by the Dirichlet conditions at the poles. The momentum sector is the expansion about the stable vacuum, with the massive linearisation \eqref{eq:chicLO}. In both cases the deviation from the vacuum is of order $1/c$, so that configurations which reach the other vacuum, such as kinks, lie outside the expansion.

\section{NR strings on $\mathbb R\times S^2$}\label{app:NC}

This section contains details of the sNC string solutions on $\mathbb R\times S^2$ used in various parts of the main text. Most solutions have appeared in literature, for example see \cite{Banerjee:2025gyh}, but we have tried to have a coherent description here anyway.

\subsection{The intrinsic sigma model}
The sigma model of non-relativistic string theory on an arbitrary string Newton-Cartan geometry
\cite{BGY} is given by
\begin{align}
\label{masteraction}
S_{P}^{NR}
={}&-\frac{T}{2}\int d^2\sigma\,\Bigg[
\sqrt{-h}\,h^{\alpha\beta}
\partial_{\alpha}X^{\mu}
\partial_{\beta}X^{\nu}H_{\mu\nu}
\nonumber\\
&\qquad\qquad
+\epsilon^{\alpha\beta}
\left(
\xi e_{\alpha}\tau_{\mu}
+\bar{\xi}\bar{e}_{\alpha}\bar{\tau}_{\mu}
\right)
\partial_{\beta}X^{\mu}
\Bigg],
\end{align}
where $h_{\alpha\beta}$ and $e_\alpha,\bar e_\alpha$ are the worldsheet metric and light-cone zweibeins, $\tau_\mu,\bar\tau_\mu=\tau_\mu^0\pm\tau_\mu^1$, and the target-space pre-contraction vielbeins expand as
\begin{equation}
e_{\mu}^{a}=c \tau^{a}_{\mu}+\frac{1}{c}m^{b}_{\mu}\epsilon^{a}_{b}+\mathcal{O}(c^{-2}),     ~~~e_{\mu}^{a'}=E_{\mu}^{a'} + \mathcal{O}(c^{-1}).
\end{equation}
Here $a = 0,1 $ denotes the longitudinal directions and $a'= 2...(D-1)$ denotes the transverse directions. One can construct metrics along the two subspaces:
\begin{equation}
H_{\mu\nu}=E_{\mu}^{a'}E_{\nu}^{b'}\delta_{a'b'}+(\tau_{\mu}^a m_{\nu}^b+\tau_{\nu}^a m_{\mu}^b)\eta_{ab}, \quad \tau_{\mu\nu} = \tau_\mu^{~a} \tau_\nu^{~b}\eta_{ab} 
\end{equation}
where  
the gauge fields $\tau_{\mu}^a$, $m_{\mu}^a$ and $E_{\mu}^{a'}$ define the string Newton-Cartan geometry.

One can also simply start from the $c$ scaled metric $ds^2=R^2(-dt^2+d\theta^2+\sin^2\theta\,d\phi_{\rm rel}^2)$, with $t,\theta$ longitudinal and $\phi_{\rm rel}$ transverse, and take
\begin{equation}\label{eq:A:limit}
R=c\,\ell,\qquad \theta=\vartheta,\qquad \phi_{\rm rel}=\frac{\varphi}{c},\qquad c\to\infty .
\end{equation}
The parent vielbeins then expand as $e^a=c\,\tau^a$ and $e^2=E^2$, with the data \eqref{eq:sncdata} for $\ell=1$. The divergent term $c^2\tau_{\mu\nu}$ of the Polyakov action is cancelled by a critical Kalb-Ramond field and replaced by Lagrange multipliers \cite{GO,BGY}. In conformal gauge, the resulting Lagrangian is \eqref{eq:NCL}, or equivalently
\begin{equation}\label{eq:A:L}
\mathcal L=-\frac T2\Big[s(\varphi'^2-\dot\varphi^2)+\xi_0(t'-\dot\vartheta)+\xi_1(\vartheta'-\dot t)\Big],
\end{equation}
with $s=\sin^2\vartheta$ and $\xi_{0,1}=\xi\pm\bar\xi$.
Varying \eqref{eq:NCL} with respect to the multipliers, to $X^\pm$ and to $\varphi$ gives
\begin{equation}\label{eq:A:eom}
\begin{gathered}
\p_-X^+=\p_+X^-=0,\qquad \p_-\xi=\p_+\bar\xi=\sin2\vartheta\,\p_+\varphi\,\p_-\varphi,\\
\p_+(s\,\p_-\varphi)+\p_-(s\,\p_+\varphi)=0 .
\end{gathered}
\end{equation}
Varying the zweibeins gives the Virasoro constraints
\begin{equation}\label{eq:A:Vir}
s(\p_+\varphi)^2+\tfrac12\xi\,\p_+X^+=0,\qquad s(\p_-\varphi)^2-\tfrac12\bar\xi\,\p_-X^-=0 ,
\end{equation}
whose relative coefficients are confirmed by their on-shell conservation. The Noether charges of $t$ and $\varphi$ are
\begin{equation}\label{eq:A:charges}
E=-\frac T2\oint\xi_1\,d\sigma,\qquad J=T\oint s\,\dot\varphi\,d\sigma .
\end{equation}
Note that the energy is an integral of a multiplier, so it is determined by the Virasoro constraints.

\subsection{Static gauge}

By \eqref{eq:A:eom}, $X^+=F(\sigma^+)$ and $X^-=G(\sigma^-)$. The static gauge $t=\kappa\tau$ then forces $F'=G'=\kappa$ and hence \eqref{eq:NCtheta}. Closure on the meridian requires $\kappa\in\mathbb Z$. The transverse equation reduces to \eqref{eq:NClin}, and \eqref{eq:A:Vir} determines the multipliers,
\begin{equation}\label{eq:A:xisol}
\xi_1=-\frac{s}{\kappa}\big(\dot\varphi^2+\varphi'^2\big),\qquad \xi_0=-\frac{2s}{\kappa}\,\dot\varphi\,\varphi' .
\end{equation}
One verifies that \eqref{eq:A:xisol} satisfies the multiplier equations in \eqref{eq:A:eom} for every solution of \eqref{eq:NClin}. Substituting into \eqref{eq:A:charges} gives the energy \eqref{eq:Eintr}.

For the rigidly rotating string $\varphi=\omega\tau$, \eqref{eq:Eintr} and \eqref{eq:A:charges} give
\begin{equation}\label{eq:A:EJ}
E=\frac{\pi T\omega^2}{2\kappa},\qquad J=\pi T\omega,\qquad E=\frac{J^2}{2\pi\kappa T},
\end{equation}
which is a Galilean, rigid-rotor dispersion relation. For the twisted rotating strings $\varphi=\omega\tau+f(\sigma)$, \eqref{eq:NClin} gives
\begin{equation}\label{eq:A:twist}
s\,f'=v,\qquad f'\simeq\frac{v}{(\vartheta-\vartheta_*)^2}
\end{equation}
near a pole $\vartheta_*$, so that closed twisted strings are singular unless $v=0$.
In terms of $x=\vartheta$ and $y=\sin x\,\varphi$, the worldsheet energy of $\varphi$ becomes
\begin{equation}\label{eq:A:Hy}
\begin{aligned}
H_\varphi=\;&\frac{T}{2\kappa}\oint d\sigma\,\Big[\dot y^2+\kappa^2\big(y_x^2-y^2\big)\Big]\\
&-\frac{T}{2}\sum_{\rm poles}\big[y^2\cot x\big].
\end{aligned}
\end{equation}
The pole terms diverge unless $y$ vanishes at every pole, so that finite energy imposes Dirichlet conditions there.

\subsection{The $1/c^2$ expansion}\label{A3}

On the background \eqref{eq:metc}, the expansion \eqref{eq:fieldexp} of the Polyakov string, with the worldsheet metric expanded similarly, gives \eqref{eq:Lexp} \cite{HH1,HH2}, with
\begin{equation}\label{eq:A:LNLO}
\begin{aligned}
\mathcal L_{\rm NLO}=\;&T\big[t_0't_1'-\dot t_0\dot t_1-\theta_0'\theta_1'+\dot\theta_0\dot\theta_1\big]\\
&-\frac T2\sin^2\theta_0\,\big(\phi_0'^2-\dot\phi_0^2\big).
\end{aligned}
\end{equation}
The variation of the subleading worldsheet metric imposes the LO Virasoro constraint, $\p_\pm X_0^+\,\p_\pm X_0^-=0$ with $X_0^\pm=t_0\pm\theta_0$. Its physical branches are
\begin{equation}\label{eq:A:branches}
\begin{aligned}
&\text{winding:} & t_0&=\kappa\tau, & \theta_0&=\kappa\sigma+\sigma_0,\\
&\text{momentum:} & t_0&=\zeta\tau+\alpha_0, & \theta_0&=\zeta\tau+\alpha_0 .
\end{aligned}
\end{equation}
On the momentum branch $\theta_0$ has no winding and the pulled-back longitudinal metric vanishes. This is the branch excluded in \cite{HH2} by the winding boundary conditions.

On the winding branch, an integration by parts brings \eqref{eq:A:LNLO} to the form \eqref{eq:NCL}, with $\vartheta=\theta_0$, $\varphi=\phi_0$ and
\begin{equation}\label{eq:A:dict}
\xi_1=2(\theta_1'-\dot t_1),\qquad \xi_0=2(\dot\theta_1-t_1').
\end{equation}
The subleading longitudinal fields are thus the intrinsic multipliers, and the energy correction is $E_{\rm NLO}=T\oint\dot t_1\,d\sigma=-\frac T2\oint\xi_1\,d\sigma$. By the NLO Virasoro constraints \eqref{eq:A:VirNLOwind} of appendix~\ref{app:levelmatch}, equivalently \eqref{eq:A:dict} combined with \eqref{eq:A:xisol}, $\dot t_1-\theta_1'=\frac{\sin^2\theta_0}{2\kappa}(\dot\phi_0^2+\phi_0'^2)$. For the rigid rotation $\phi_0=\omega\tau$, we write $t_1=h\tau$ plus terms periodic in $\sigma$ and average over $\sigma$. Since $\theta_1$ is periodic, $\oint\theta_1'\,d\sigma=0$, and we obtain $h=\frac{\omega^2}{2\kappa}\langle\sin^2\theta_0\rangle=\frac{\omega^2}{4\kappa}$. Hence $t_1=\frac{\omega^2}{4\kappa}\tau$ up to periodic terms, so that the corrected static-gauge frequency is
\begin{equation}\label{eq:A:kr}
\kappa_r=\kappa+\frac{\omega^2}{4\kappa\,c^2}+O(c^{-4}).
\end{equation}

On the momentum branch, the NLO transverse equation is
\begin{equation}\label{eq:A:momeom}
\p_\tau\big(\sin^2\theta_0\,\dot\phi_0\big)=\p_\sigma\big(\sin^2\theta_0\,\phi_0'\big),
\end{equation}
and the NLO Virasoro constraints determine $X_1^-=t_1-\theta_1$ algebraically,
\begin{equation}\label{eq:A:X1m}
\p_\tau X_1^-=\frac{\sin^2\theta_0}{2\zeta}\big(\dot\phi_0^2+\phi_0'^2\big),\qquad \p_\sigma X_1^-=\frac{\sin^2\theta_0}{\zeta}\,\dot\phi_0\,\phi_0' ,
\end{equation}
as in light-cone gauge. 

\subsection{The pulsating string}\label{app:puls}
 
On the sphere of radius $c$, in static gauge $t=\zeta_r\tau$, the pulsating string is $\phi_{\rm rel}=(m/c)\sigma$ with $\theta=\theta(\tau)$ given by \eqref{eq:pulsexact}. Its energy, and its oscillation number, i.e.\ the action variable of the motion in $\theta$, are exactly
\begin{equation}\label{eq:A:pulsN}
E=2\pi Tc^2\zeta_r,\qquad N=\frac1{2\pi}\oint P_\theta\,d\theta=4Tc^2\zeta_r\,\mathrm E(k_c^2),
\end{equation}
with $P_\theta=2\pi Tc^2\dot\theta$, $k_c=m/(c\zeta_r)$, and $\mathrm E$ the complete elliptic integral of the second kind. The momentum $P_\theta$ is not conserved, since $\theta$ is not an isometry, which is why $N$ is used instead. Holding $N=c^2N_{\rm LO}$ fixed, with $N_{\rm LO}=2\pi T\zeta$, requires $\zeta_r=\zeta+m^2/(4\zeta c^2)+O(c^{-4})$, the counterpart of \eqref{eq:A:kr}, and gives
\begin{equation}\label{eq:A:puls}
N=c^2N_{\rm LO},\qquad E=c^2N_{\rm LO}+\frac{(\pi Tm)^2}{N_{\rm LO}}+O(c^{-2}),
\end{equation}
which we have checked numerically against \eqref{eq:A:pulsN}. The first term is the LO energy of the point-like string, and the second the NLO energy of the transverse winding. Equivalently, the light-cone energy is
\begin{equation}\label{eq:A:lc}
\begin{aligned}
E-P_\theta\Big|_{\rm NLO}&=\frac{1}{2p^+}\oint T^2\sin^2\theta_0\,(\phi_0')^2\,d\sigma\\
&=\frac{2\pi^2T^2m^2}{N_{\rm LO}}\sin^2\theta_0 ,
\end{aligned}
\end{equation}
with $p^+=N_{\rm LO}/(2\pi)$. Its time average reproduces the NLO energy of \eqref{eq:A:puls}, the remainder being the oscillating part of $P_\theta$.

\subsection{Zero modes of the NLO Virasoro constraints}\label{app:levelmatch}

In conformal gauge, the relativistic Virasoro constraints on \eqref{eq:metc} are
\begin{equation}\label{eq:A:Virrel}
c^2\Big[-(\p_\pm t)^2+(\p_\pm\theta)^2\Big]+\sin^2\theta\,(\p_\pm\phi)^2=0 .
\end{equation}
Expanding via \eqref{eq:fieldexp} gives the LO constraints $\p_\pm X_0^+\,\p_\pm X_0^-=0$ at $O(c^2)$ and the NLO constraints at $O(c^0)$,
\begin{equation}\label{eq:A:VirNLO}
2\big(\p_\pm\theta_0\,\p_\pm\theta_1-\p_\pm t_0\,\p_\pm t_1\big)+\sin^2\theta_0\,(\p_\pm\phi_0)^2=0 .
\end{equation}
With $s=\sin^2\theta_0$, these determine $(t_1,\theta_1)$ locally in terms of $\phi_0$. On the winding branch ($\p_\pm t_0=\frac\kappa2$, $\p_\pm\theta_0=\pm\frac\kappa2$), taking their sum and difference yields
\begin{equation}\label{eq:A:VirNLOwind}
\dot t_1-\theta_1'=\frac{s}{2\kappa}\big(\dot\phi_0^2+\phi_0'^2\big),\qquad
\dot\theta_1-t_1'=-\frac{s}{\kappa}\,\dot\phi_0\,\phi_0' ,
\end{equation}
which match the intrinsic multiplier relations \eqref{eq:A:xisol} under \eqref{eq:A:dict}. On the momentum branch ($\p_\pm t_0=\p_\pm\theta_0=\frac\zeta2$), \eqref{eq:A:VirNLO} becomes $\p_\pm X_1^-=\frac{s}{\zeta}(\p_\pm\phi_0)^2$ for $X_1^-=t_1-\theta_1$, reproducing \eqref{eq:A:X1m}.

On a closed string, periodicity in $\sigma$ imposes global zero-mode conditions that differ between the two branches. On the winding branch, $t_1-h\tau$ is periodic, so integrating the second relation of \eqref{eq:A:VirNLOwind} over $\sigma\in[0,2\pi)$ gives
\begin{equation}\label{eq:A:drift}
\frac{d}{d\tau}\oint\theta_1\,d\sigma=-\frac1\kappa\oint\sin^2\theta_0\,\dot\phi_0\,\phi_0'\,d\sigma .
\end{equation}
Rather than constraining $\phi_0$, \eqref{eq:A:drift} fixes the drift rate of the $O(c^{-2})$ meridian zero mode $\oint\theta_1\,d\sigma$ (which vanishes for the rigid rotation $\phi_0'=0$). Because $\theta$ is not an isometry and $\sin^2\theta_0$ depends on $\sigma$, neither side is conserved, unlike on an isometric circle where this zero mode gives standard level-matching.

On the momentum branch, $\theta_0=\zeta\tau+\alpha_0$ has no winding and $X_1^-=t_1-\theta_1$ is periodic. Integrating $\p_\sigma X_1^-$ in \eqref{eq:A:X1m} over the closed string and factoring out $\sin^2\theta_0(\tau)$ yields a genuine level-matching constraint on $\phi_0$,
\begin{equation}\label{eq:A:levelmatch}
\oint\dot\phi_0\,\phi_0'\,d\sigma=0 .
\end{equation}
Since $\sin^2\theta_0$ is independent of $\sigma$, \eqref{eq:A:momeom} ensures $\frac{d}{d\tau}\oint\sin^2\theta_0\,\dot\phi_0\,\phi_0'\,d\sigma=0$, so \eqref{eq:A:levelmatch} is preserved in time. It holds trivially for the pulsating string ($\phi_0=m\sigma$, $\dot\phi_0=0$) and requires equal left- and right-moving transverse worldsheet momenta in general.

\subsection{Solutions outside the intrinsic theory}\label{app:larger}

Let us make precise in which sense the solution space of the expansion is larger than that of the intrinsic theory. On the winding branch of \eqref{eq:A:branches}, the NLO Lagrangian \eqref{eq:A:LNLO} coincides with the intrinsic one through \eqref{eq:A:dict}, so that its solutions are the intrinsic solutions, up to the residual reparametrisations. The branch obtained from it by worldsheet parity, $\sigma\to-\sigma$, contains the same configurations. On the momentum branch, by contrast,
\begin{equation}\label{eq:A:mombranch}
X_0^+=2(\zeta\tau+\alpha_0),\qquad X_0^-=0,
\end{equation}
which satisfies the LO Virasoro constraint, but violates the intrinsic chirality constraint $\p_-X^+=0$, since
\begin{equation}\label{eq:A:chiralviol}
\p_-X_0^+=\zeta\neq0 .
\end{equation}
Equivalently, the pulled-back longitudinal metric $\tau(\p_+X,\p_-X)=-\frac12\big(\p_+X^+\p_-X^-+\p_-X^+\p_+X^-\big)$ vanishes identically on \eqref{eq:A:mombranch}. The momentum sector, and in particular the pulsating string \eqref{eq:A:puls}, therefore consists of solutions of the expansion with no counterpart in the intrinsic theory. This is the branch excluded in \cite{HH2} by the assumption of a Lorentzian pull-back.

\section{Relativistic analogue of the reduced field}\label{app:chirel}

In the relativistic theory on $\mathbb R\times S^2$ of radius $c$, define the vielbeins
\begin{equation}
e^0=c\,dt,\qquad e^1=c\,d\theta,\qquad e^2=\sin\theta\,d\phi,
\end{equation}
so that $ds^2=-(e^0)^2+(e^1)^2+(e^2)^2$. The relativistic analogue of the sNC invariant \eqref{eq:chicov} is the normalised oriented area
\begin{equation}\label{eq:chirel}
\chi_{\rm rel}
\equiv
\frac{(e^1\wedge e^2)(\p_+X,\p_-X)}
{e^0(\p_+X)\,e^0(\p_-X)} .
\end{equation}
In components,
\begin{equation}\label{eq:chirelcomp}
\chi_{\rm rel}
=
\frac{\sin\theta\,(\p_+\theta\,\p_-\phi-\p_-\theta\,\p_+\phi)}
{c\,\p_+t\,\p_-t} .
\end{equation}
In static gauge $t=\kappa_r\tau$ this becomes
\begin{equation}\label{eq:chirelstatic}
\chi_{\rm rel}
=
\frac{4\sin\theta\,(\p_+\theta\,\p_-\phi-\p_-\theta\,\p_+\phi)}
{c\,\kappa_r^2}
=
\frac{2\sin\theta\,(\theta'\dot\phi-\dot\theta\phi')}
{c\,\kappa_r^2} .
\end{equation}

The Pohlmeyer field is defined by \eqref{eq:psidef}
with $f'_\pm=c\kappa_r/2$ in static gauge. Expanding \eqref{eq:chirelstatic} on the winding branch, where $\theta=\kappa\sigma+\sigma_0+O(c^{-2})$ and $\phi=\varphi+O(c^{-2})$, gives
\begin{equation}\label{eq:chirelexp}
\chi_{\rm rel}
=
\frac{2\chi}{c}+O(c^{-3}),
\qquad
\chi=\frac{\sin\vartheta\,\dot\varphi}{\kappa}.
\end{equation}
On the same branch, the expansion of \eqref{eq:psidef} near $\psi=\pi$ gives
\begin{equation}\label{eq:psiexp}
\psi=\pi-\frac{2\chi}{c}+O(c^{-3}).
\end{equation}
Combining \eqref{eq:chirelexp} and \eqref{eq:psiexp},
\begin{equation}\label{eq:chirelpsi}
\chi_{\rm rel}=\pi-\psi+O(c^{-3}),
\qquad
\psi=\pi-\chi_{\rm rel}+O(c^{-3}).
\end{equation}
Thus $\chi_{\rm rel}$ is the deviation of the Pohlmeyer angle from the unstable vacuum $\psi=\pi$, and the sNC field $\chi$ is its leading coefficient:
\begin{equation}\label{eq:chisnc}
\chi=\frac{c}{2}(\pi-\psi)+O(c^{-2}).
\end{equation}
On the momentum branch, where $\psi\to0$, the same definition gives $\chi_{\rm rel}=-2\chi_c/c+O(c^{-3})$, which is the relation $\chi_{\rm sNC}=-\chi_c$ of section~\ref{sec:expansion}.

\begin{figure*}[t]
\centering
\begin{tikzpicture}[>={Stealth[length=2.2mm]},scale=1.0]
  \node[font=\small\bfseries] at (-0.75,2.75) {(a)};
  \begin{scope}
    \node[font=\scriptsize] at (1.6,2.75) {generic $a\notin\mathbb Z$};
    \draw[->,gray!60] (-0.3,0) -- (3.4,0) node[right,font=\scriptsize,black] {$P_1$};
    \draw[->,gray!60] (0,-0.3) -- (0,2.4) node[left,font=\scriptsize,black] {$P_2$};
    \coordinate (C) at (1.85,1.05);
    \draw[dashed,blue!60!black,thin] (C) circle (0.95);
    \fill[blue!70!black] (C) circle (1.4pt) node[right,font=\scriptsize] {$C$};
    \coordinate (S) at ($(C)+(210:0.95)$);
    \coordinate (E) at ($(C)+(100:0.95)$);
    \draw[->,very thick,blue!70!black] ($(C)+(212:1.15)$) arc (212:102:1.15);
    \node[blue!70!black,font=\scriptsize] at ($(C)+(160:1.45)$) {$2\pi a$};
    \fill (S) circle (1.5pt) node[below left=-1pt,font=\scriptsize] {$\sigma=0$};
    \fill (E) circle (1.5pt) node[above right=-1pt,font=\scriptsize] {$\sigma=2\pi$};
    \draw[->,thick,green!50!black] (S) -- ++(0:0.42);
    \draw[->,thick,green!50!black] (S) -- ++(90:0.42);
    \draw[->,thick,green!50!black] (E) -- ++(-110:0.42);
    \draw[->,thick,green!50!black] (E) -- ++(-20:0.42);
    \node[font=\scriptsize,align=center] at (1.6,-0.7) {rotation about $C$:\\ translation removable};
  \end{scope}
  \begin{scope}[xshift=4.4cm]
    \node[font=\scriptsize] at (1.6,2.75) {resonant $a=n\kappa$};
    \draw[->,gray!60] (-0.3,0) -- (3.4,0) node[right,font=\scriptsize,black] {$P_1$};
    \draw[->,gray!60] (0,-0.3) -- (0,2.4) node[left,font=\scriptsize,black] {$P_2$};
    \coordinate (S2) at (0.5,0.45);
    \coordinate (E2) at (2.55,1.55);
    \draw[->,very thick,red!70!black] (S2) -- (E2) node[midway,below right=-1pt,font=\scriptsize] {$\vec\Omega$};
    \fill (S2) circle (1.5pt) node[below left=-1pt,font=\scriptsize] {$\sigma=0$};
    \fill (E2) circle (1.5pt) node[below right=0pt,font=\scriptsize] {$\sigma=2\pi$};
    \draw[->,thick,green!50!black] (S2) -- ++(0.42,0);
    \draw[->,thick,green!50!black] (S2) -- ++(0,0.42);
    \draw[->,thick,green!50!black] (E2) -- ++(0.42,0);
    \draw[->,thick,green!50!black] (E2) -- ++(0,0.42);
    \draw[->,blue!70!black,thick] (0.95,1.75) arc (0:330:0.2);
    \node[blue!70!black,font=\scriptsize,right] at (1.0,1.78) {$e^{2\pi aJ}=\mathbf 1$};
    \node[font=\scriptsize,align=center] at (1.6,-0.7) {pure translation:\\ $\mathcal Q=|\vec\Omega|^2$ invariant};
  \end{scope}
  \begin{scope}[xshift=10.4cm]
    \node[font=\small\bfseries] at (-0.75,2.75) {(b)};
    \draw[->,gray!60] (-0.2,0) -- (3.8,0) node[right,font=\scriptsize,black] {$\sigma$};
    \draw[->,gray!60] (0,-0.3) -- (0,2.5) node[left,font=\scriptsize,black] {$\tau$};
    \foreach \x in {0.55,1.2,1.85,2.5} {
      \draw[green!50!black,thin] (\x,0) -- ++(1.9/1.155,1.9);
    }
    \node[green!50!black,font=\scriptsize,align=left,anchor=west] at (3.55,1.25) {constant\\phase};
    \draw[->,very thick,blue!70!black] (0,0) -- (3.1,0);
    \node[blue!70!black,font=\scriptsize,below] at (1.55,-0.02) {$\lambda=1$: $M(1)=\mathbf 1$};
    \draw[->,very thick,red!70!black] (0,0) -- (2.5,2.165);
    \node[red!70!black,font=\scriptsize,above left=-1pt] at (2.5,2.165) {$\lambda=\lambda_n^\mp$};
    \draw[red!70!black,thin] (0.7,0) arc (0:40.9:0.7);
    \node[red!70!black,font=\scriptsize] at (0.95,0.3) {$\theta_n$};
    \node[font=\scriptsize,align=center] at (1.8,-0.75) {slope $\tanh\theta_n=\omega_n/p_n$:\\ the mode is stationary on this slice};
  \end{scope}
\end{tikzpicture}
\caption{Physical meaning of the monodromy. (a) Transverse holonomy of the moving frame in the $(P_1,P_2)$ plane. For generic $a\notin\mathbb Z$ (left), the frame returns rotated by $2\pi a$, and the combination of this rotation with a translation is a pure rotation about a shifted centre $C$, so that the translation can be removed and no invariant remains. At a resonant value $a=n\kappa$ (right), the frame returns with its original orientation, $e^{2\pi aJ}=\mathbf 1$, displaced by the translation $\vec\Omega=(\Omega_1,\Omega_2)$, whose length squared $\mathcal Q$ is invariant. (b) The resonant spectral parameters on the worldsheet, shown for $n=2$. The lines of constant phase of the $n$-th right- or left-moving mode (green) have slope $p_n/\omega_n=\coth\theta_n$, and are the worldlines of its co-moving frame. The monodromy at $\lambda=\lambda_n^\mp$ transports the frame along a slice tilted by the rapidity $\theta_n$, with slope $\tanh\theta_n=\omega_n/p_n$, which is a line of constant co-moving time. Along this slice the mode is stationary, with wavenumber $\kappa$. On the slice $\lambda=1$, the frame closes on itself.}
\label{fig:monodromy}
\end{figure*}
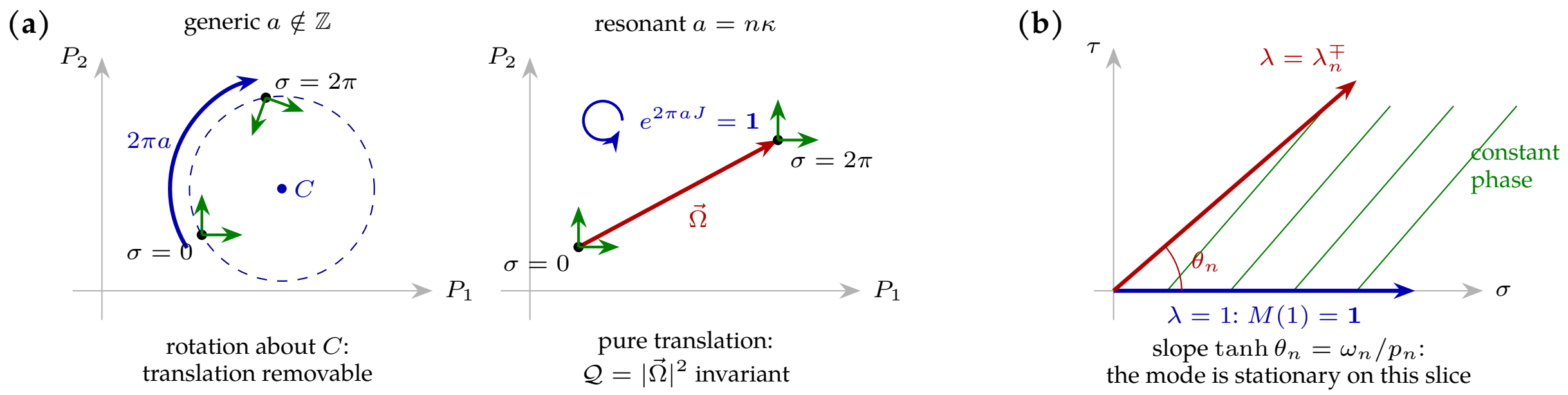

\section{Physical meaning of the monodromy}\label{sec:monophys}

The structure of the monodromy \eqref{eq:monodromy}, and in particular the fact that conserved charges arise only at the discrete spectral parameters \eqref{eq:resonance}, admits a direct physical interpretation from two complementary viewpoints: the geometry of the moving frame, and worldsheet kinematics (figure~\ref{fig:monodromy}).

\paragraph{Transverse holonomy on a boosted slice.}
At $\lambda=1$, the connection \eqref{eq:LaxNC} is the Maurer-Cartan form \eqref{eq:MCresult} of the moving frame $\mathcal F$ in \eqref{eq:frame}, so that $A_\sigma=\p_\sigma\mathcal F\,\mathcal F^{-1}$. The Wilson line is therefore integrated explicitly as $U(\sigma;1)=\mathcal F(\sigma)\mathcal F(0)^{-1}$. Because the string is closed, $\vartheta(2\pi)=\vartheta(0)+2\pi\kappa$ with $\kappa\in\mathbb Z$ and the transverse displacement $y(\sigma)$ is periodic, so that $\mathcal F(2\pi)=\mathcal F(0)$ and
\begin{equation}\label{eq:M1trivial}
M(1)=\mathbf 1
\end{equation}
identically for every closed-string configuration. Around a constant-$\tau$ slice of the worldsheet, the moving frame closes on itself with no net transverse mismatch. This is the geometric reason why the rigid rotation ($n=1$, for which $\lambda_1^\pm=1$) carries no monodromy charge: at any fixed instant, a rigidly rotating meridian closes seamlessly at the poles.

When $\lambda=e^{-\rho}\neq1$, the boost relation \eqref{eq:boostpull} shows that $M(\lambda)$ parallel-transports the frame along a worldsheet slice tilted in time by rapidity $\rho$. Because of worldsheet time dilation, as $\sigma$ runs from $0$ to $2\pi$ along this tilted slice, the longitudinal angle advances by
\begin{equation}\label{eq:DeltaTheta}
\Delta\vartheta_{\rm boost}=2\pi a=2\pi\kappa\cosh\rho .
\end{equation}
For generic $\rho$, $a$ is not an integer, and the longitudinal frame returns rotated by a non-trivial angle $2\pi a\notin 2\pi\mathbb Z$. In the Euclidean plane, a rotation accompanied by a translation $(\Omega_1,\Omega_2)$ is simply a pure rotation about a shifted origin (figure~\ref{fig:monodromy}a). The translational part of $M(\lambda)$ can therefore be completely removed by a global shift of the transverse origin, leaving no gauge-invariant information about the string profile.

At the discrete rapidities $\rho=\mp\theta_n$ where $a=n\kappa\in\mathbb Z$, by contrast, the boosted longitudinal frame completes an integer number $n\kappa$ of full turns around the meridian and returns to its exact starting orientation, $e^{2\pi aJ}=\mathbf 1$. Because the slice is tilted across a time interval $\Delta\tau\neq0$, however, the vibrating string moves transversely during the circuit and does not return to its initial transverse state. The frame therefore fails to close by a pure transverse translation $\vec\Omega=(\Omega_1,\Omega_2)$ (figure~\ref{fig:monodromy}a). With no net rotation, this mismatch cannot be shifted away by changing the origin: it is a genuine transverse holonomy defect, and its length squared $\mathcal Q=\Omega_1^2+\Omega_2^2$ is invariant under all $e(2)$ frame transformations.

\paragraph{Co-moving resonance with tachyonic waves.}
Why does the $n$-th resonant slice isolate only the $n$-th oscillator mode? By \eqref{eq:modeLR}, the $n$-th mode has worldsheet wavenumber $p_n=n\kappa$ and frequency $\omega_n=\kappa\sqrt{n^2-1}$, which lie on the spacelike mass-shell hyperboloid
\begin{equation}\label{eq:tachshell}
p_n^2-\omega_n^2=\kappa^2 .
\end{equation}
Comparing \eqref{eq:tachshell} with the resonance condition $\cosh\theta_n=n$ in \eqref{eq:resonance}, we see that the worldsheet two-momentum of the $n$-th mode is precisely
\begin{equation}\label{eq:momrapidity}
(p_n,\omega_n)=\kappa(\cosh\theta_n,\sinh\theta_n).
\end{equation}
The resonant spectral parameters $\lambda_n^\pm=e^{\pm\theta_n}$ are therefore the exact worldsheet rapidities of the tachyonic modes.

Because the reduced equation \eqref{eq:chiLO} is tachyonic, the phase velocity of its modes on the worldsheet is subluminal,
\begin{equation}\label{eq:vphase}
v_{{\rm ph},n}=\frac{\omega_n}{p_n}=\tanh\theta_n<1,\qquad n>1 .
\end{equation}
Unlike ordinary massive modes (whose phase velocity is superluminal), each tachyonic mode with $n>1$ admits a co-moving worldsheet frame, reached by boosting with rapidity $\pm\theta_n$, in which its frequency vanishes ($\tilde\omega_n=0$) and its wavenumber is Lorentz-contracted to the fundamental meridian winding, $\tilde p_n=\kappa$ (figure~\ref{fig:monodromy}b). Setting $\lambda=\lambda_n^\mp$ boosts the Lax connection into the co-moving frame of the $n$-th right- or left-moving wave. In this frame, the $n$-th wave is stationary and its spatial period matches the rotation rate $\kappa$ of the longitudinal meridian frame. Its transverse velocity and gradient therefore add up coherently along the string in \eqref{eq:Omega}, whereas all other modes $m\neq n$ oscillate out of phase with the frame and average to zero.

\section{On NNLO Lax pairs}\label{NNLOlax}

 To exhibit the NNLO structure, we expand the relativistic pair \eqref{eq:Lax}
strictly order by order, using
\begin{equation}\label{eq:strictdata}
\kappa_r=\kappa+\frac{h}{c^2}+O(c^{-4}),\qquad
\chi=\chi_{\rm NLO}+\frac{\chi_{\rm NNLO}}{c^2}+O(c^{-4}),
\end{equation}
with $\psi=\pi-2\chi/c$ and $\mu=\kappa_r$. The components are
\begin{equation}\label{eq:strictA}
\begin{aligned}
A_+^{(0)}&=\frac{\lambda\kappa}{4}\sigma_1,
& A_+^{(1)}&=i\,\p_+\chi_{\rm NLO}\,\sigma_3,\\
A_+^{(2)}&=\frac{\lambda h}{4}\sigma_1,
& A_+^{(3)}&=i\,\p_+\chi_{\rm NNLO}\,\sigma_3,
\end{aligned}
\end{equation}
and
\begin{equation}\label{eq:strictAminus}
\begin{aligned}
A_-^{(0)}&=\frac{\kappa}{4\lambda}\sigma_1, \quad
A_-^{(1)}=-\frac{\kappa}{2\lambda}\chi_{\rm NLO}\,\sigma_2,\\
A_-^{(2)}&=\frac{1}{4\lambda}\big(h-2\kappa\chi_{\rm NLO}^2\big)\sigma_1,\\
A_-^{(3)}&=-\frac{1}{2\lambda}\Big(h\chi_{\rm NLO}+\kappa\chi_{\rm NNLO}
-\frac{2\kappa}{3}\chi_{\rm NLO}^3\Big)\sigma_2 .
\end{aligned}
\end{equation}
The parity pattern noted below \eqref{eq:Laxwind} is manifest:
$A_\pm^{(n)}$ lies along $\sigma_1$ for even $n$ and along
$\sigma_{2,3}$ for odd $n$. The exact curvature is odd in $1/c$, so
$F^{(2k)}=0$ identically and the reduced equations come from
$F^{(2k+1)}$. At order $c^{-3}$,
\begin{equation}\label{eq:F3sum}
\begin{aligned}
F^{(3)}={}&\p_-A_+^{(3)}-\p_+A_-^{(3)}
+\big[A_+^{(0)},A_-^{(3)}\big]\\
&+\big[A_+^{(1)},A_-^{(2)}\big]
+\big[A_+^{(2)},A_-^{(1)}\big]\\
&+\big[A_+^{(3)},A_-^{(0)}\big].
\end{aligned}
\end{equation}
The $\sigma_3$ component receives contributions from three terms:
\begin{equation}\label{eq:F3sigma3}
\begin{aligned}
\p_-A_+^{(3)}
&=i\,\p_+\p_-\chi_{\rm NNLO}\,\sigma_3,\\
\big[A_+^{(0)},A_-^{(3)}\big]
&=-\frac{i\kappa}{4}\Big(h\chi_{\rm NLO}+\kappa\chi_{\rm NNLO}-\frac{2\kappa}{3}\chi_{\rm NLO}^3\Big)\sigma_3,\\
\big[A_+^{(2)},A_-^{(1)}\big]
&=-\frac{i\kappa h}{4}\chi_{\rm NLO}\,\sigma_3,
\end{aligned}
\end{equation}
while $-\p_+A_-^{(3)}$, $[A_+^{(1)},A_-^{(2)}]$ and
$[A_+^{(3)},A_-^{(0)}]$ lie along $\sigma_2$ and cancel among
themselves. Hence
\begin{equation}\label{eq:F3result}
\begin{aligned}
F^{(3)}=i\Big[{}&\p_+\p_-\chi_{\rm NNLO}
-\frac{\kappa^2}{4}\chi_{\rm NNLO}\\
&-\frac{\kappa h}{2}\chi_{\rm NLO}
+\frac{\kappa^2}{6}\chi_{\rm NLO}^3\Big]\sigma_3 .
\end{aligned}
\end{equation}
Flatness of \eqref{eq:Lax} at order $c^{-3}$ is therefore equivalent
to \eqref{eq:chiNNLOeq}, including its cubic term. Note where the cubic term originates: it comes from the
commutator $\big[A_+^{(0)},A_-^{(3)}\big]$ of the LO rotation with the
$\chi_{\rm NLO}^3$ part of $A_-^{(3)}$, which is part of the covariant
derivative $D^{(0)}A^{(3)}$, and not from the commutators of
$A^{(1)}$ with $A^{(2)}$. The latter are present and necessary, but
they contribute to the $\sigma_2$ sector, which cancels.\footnote{Charges
sensitive to the cubic coupling of \eqref{eq:chiexact} would require the
monodromy of the pair \eqref{eq:strictA}--\eqref{eq:strictAminus},
whose algebra is no longer the contraction $e(2)$. We have not
analysed it.}

The contraction $e(2)$ is thus exact only through NLO. At order
$c^{-2}$ the rotation generator $\sigma_1$ acquires a correction
quadratic in $\chi_{\rm NLO}$, cf.\ $A_-^{(2)}$ in
\eqref{eq:strictAminus}, and the commutator
$[\sigma_2,\sigma_3]\propto\sigma_1$ of the two would-be translations,
which vanishes in the contraction, re-enters. The full $su(2)$
structure is restored order by order in $1/c^2$. A zero-curvature
representation of a non-linear equation is not automatic. Here it is
inherited from the integrability of the sine-Gordon equation at finite
$c$. We have not established the relation between $\chi$ and the string
fields beyond NLO.

\bibliographystyle{unsrt}
\footnotesize\setlength{\itemsep}{0pt}
\bibliography{combined_references_arxiv}\end{document}